\documentclass[journal=chreay,manuscript=article]{achemso}
\usepackage{graphicx,subcaption}
\usepackage{color}
\usepackage{hyperref}
\usepackage{float}
\usepackage{array, multirow}
\usepackage{amsmath}
\usepackage{amsfonts}
\usepackage{newtxtext}
\usepackage{booktabs}
\usepackage{xcolor}
\usepackage{soul}
\usepackage{url}
\usepackage{physics}

\hypersetup{colorlinks,
    linkcolor={blue!75!black!80!yellow},
    citecolor={blue!75!black!80!yellow},
    urlcolor={blue!75!black!80!yellow}
    }
\title{\large Quantum Algorithms and Applications for Open Quantum Systems}

\author{Luis H. Delgado-Granados}
\affiliation{Department of Chemistry and The James Franck Institute, The University of Chicago, Chicago, IL 60637 USA}
\author{Timothy J. Krogmeier}
\affiliation{Department of Chemistry, Washington University in St. Louis, St. Louis, MO 61630 USA}
\alsoaffiliation{Department of Chemistry, University of Minnesota, Minneapolis, MN 55455 USA}
\author{LeeAnn M. Sager-Smith}
\affiliation{Department of Chemistry and Physics, Saint Mary's College, Notre Dame, IN 46556 USA}
\author{Irma Avdic}
\affiliation{Department of Chemistry and The James Franck Institute, The University of Chicago, Chicago, IL 60637 USA}
\author{Zixuan Hu}
\affiliation{Department of Chemistry, Department of Electrical and Computer Engineering, and Purdue Quantum Science and Engineering Institute, Purdue University, West Lafayette, IN 47907, USA}
\alsoaffiliation{Key Laboratory of Precision and Intelligent Chemistry, University of Science and Technology of China, Hefei 230026, China}
\author{Manas Sajjan}
\affiliation{Department of Chemistry, Department of Electrical and Computer Engineering, and Purdue Quantum Science and Engineering Institute, Purdue University, West Lafayette, IN 47907, USA}
\author{Maryam Abbasi}
\affiliation{Department of Chemistry, Washington University in St. Louis, St. Louis, MO 61630 USA}
\alsoaffiliation{Department of Chemistry, University of Minnesota, Minneapolis, MN 55455 USA}
\author{Scott E. Smart}
\affiliation{Division of Physical Sciences, College of Letters and Science,
University of California, Los Angeles, CA 90095, USA}
\author{Prineha Narang}
\affiliation{Division of Physical Sciences, College of Letters and Science,
University of California, Los Angeles, CA 90095, USA}
\author{Sabre Kais}
\affiliation{Department of Chemistry, Department of Electrical and Computer Engineering, and Purdue Quantum Science and Engineering Institute, Purdue University, West Lafayette, IN 47907, USA}
\affiliation{Department of Chemistry, Department of Electrical and Computer Engineering, and Purdue Quantum Science and Engineering Institute, Purdue University, West Lafayette, IN 47907, USA}
\author{Anthony W. Schlimgen}
\affiliation{Department of Chemistry, Washington University in St. Louis, St. Louis, MO 61630 USA}
\alsoaffiliation{Department of Chemistry, University of Minnesota, Minneapolis, MN 55455 USA}
\author{Kade Head-Marsden}
\email{khm@umn.edu}
\affiliation{Department of Chemistry, Washington University in St. Louis, St. Louis, MO 61630 USA}
\alsoaffiliation{Department of Chemistry, University of Minnesota, Minneapolis, MN 55455 USA}
\author{David A. Mazziotti}
\email{damazz@uchicago.edu}
\affiliation{Department of Chemistry and The James Franck Institute, The University of Chicago, Chicago, IL 60637 USA}

\date{Submitted June 5, 2024}

\begin{document}

\begin{abstract}
Accurate models for open quantum systems\textemdash quantum states that have non-trivial interactions with their environment\textemdash may aid in the advancement of a diverse array of fields, including quantum computation, informatics, and the prediction of static and dynamic molecular properties. In recent years, quantum algorithms have been leveraged for the computation of open quantum systems as the predicted quantum advantage of quantum devices over classical ones may allow previously inaccessible applications. Accomplishing this goal will require input and expertise from different research perspectives, as well as the training of a diverse quantum workforce, making a compilation of current quantum methods for treating open quantum systems both useful and timely. In this Review, we first provide a succinct summary of the fundamental theory of open quantum systems and then delve into a discussion on recent quantum algorithms. We conclude with a discussion of pertinent applications, demonstrating the applicability of this field to realistic chemical, biological, and material systems.
\end{abstract}

\maketitle

\section{Introduction}

Open quantum systems (OQS) arise when a quantum system has non-negligible interactions with its environment. Theoretical treatment of an open quantum system entails separating the degrees of freedom into a \emph{system} and the remaining degrees of freedom as an \emph{environment} or \emph{bath}.~\cite{Breuer2007,Rivas2012} The dynamics of OQS are often divided into two primary categories. The first category, which we call \emph{Markovian}, is when the relaxation time of the system is significantly slower than that of the environment, and second, which we call \emph{non-Markovian}, is when the time scales of relaxation within the system and the environment are comparable~\cite{Breuer2007, Rivas2012,Breuer2016,Rivas2014, Li2019a, Li2019b}. OQS have potentially significant applications in advancing quantum computing~\cite{Kraus2008,Diehl2008,Verstraete2009,HeadMarsdenFlick2021,Olivera-Atencio2023}, cryptography~\cite{Gisin2002}, metrology~\cite{Kolkowitz2016,Martin2013}, simulation~\cite{Georgescu2014}, thermodynamics~\cite{YungerHalpern2016}, control~\cite{Wiseman2009,Pechen2014,Wu2007}, and information processing,~\cite{Avdic2023b} as well as in understanding and optimizing chemical reactions in solvents, thereby bridging fundamental science with practical applications across technology, biology~\cite{Mohseni2014,Mohseni2013}, and materials science~\cite{Kitai2020}.

 Modeling OQS on quantum computers, in particular, has the potential to revolutionize various classes of computation by harnessing the unique information processing available in quantum mechanics~\cite{Nielsen2010,Wilde2013,Kassal2011}. While classical computers perform their calculations through combinations of bits, with each bit being a Boolean choice of 0 or 1, quantum computers perform their calculations through combinations of quantum bits, or qubits, in which each qubit can be any linear superposition of 0 and 1. A crucial application area for quantum computing is the prediction of the static and dynamic properties of many-electron atoms, molecules, and materials. Although a plethora of algorithms have been proposed,~\cite{Cao2019, Tilly2022, Miessen2023, Fauseweh2024, Cerezo2021, Bharti2022, Wecker2015} most algorithms have focused on treating electronic structure in closed quantum systems. In this review article, we summarize a new, emerging frontier of quantum algorithms for open quantum systems. Because these systems require significant computational resources on classical devices, their simulation on quantum devices has the potential for realizing significant computational savings and reaching important, potentially hitherto inaccessible applications.

\begin{figure}[h!]
    \centering
    \includegraphics[width = 0.85\textwidth]{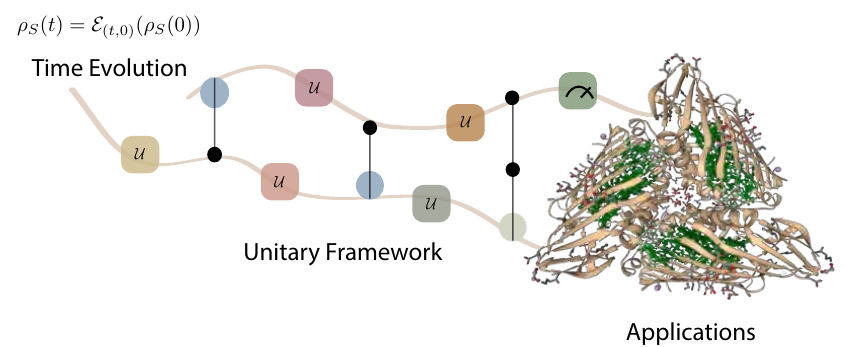}
    \caption{Schematic representation of the translation of OQS dynamics into a quantum computing framework for potential applications beyond classical capabilities.}
    \label{fig:overview}
\end{figure}

\begin{figure}[h!]
    \centering
    \includegraphics[width = 0.5\textwidth]{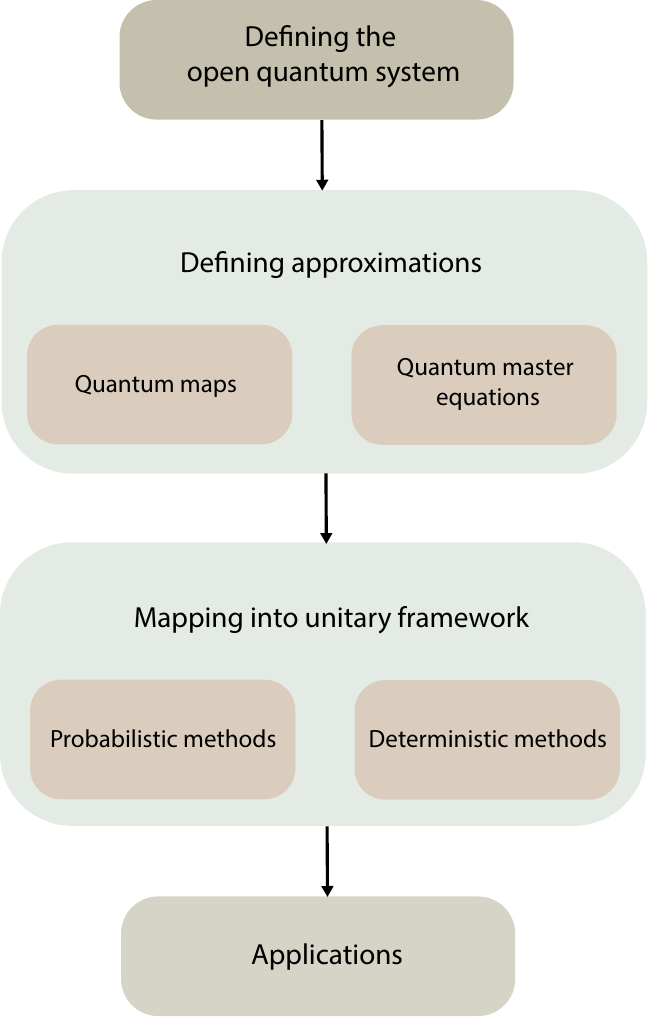}
    \caption{Step-by-step representation of the mapping from a classical open system approach into a quantum open system framework.}
    \label{fig:outline}
\end{figure}

 In recent years, with the rapid progress in different quantum computing platforms such as superconducting qubits~\cite{Huang2020}, ion-traps~\cite{Bruzewicz2019}, and cold atoms~\cite{Wintersperger2023}, it has become desirable to test and benchmark new algorithms on quantum devices. To this end, a variety of approaches have been proposed, benchmarked, and applied to model OQS using quantum computers~\cite{Bacon2001, Wang2011, Sweke2015,Sweke2016,Wei2016, Hu2020, Patsch2020,Garcia-Perez2020,Endo2020,Schlimgen2021,Kamakari2022, Suri2023}. Because quantum computers work within the language of unitary transformations~\cite{Nielsen2010,Wilde2013}, a fundamental challenge to simulating OQS on quantum devices is to treat their \emph{non-unitary} dynamics. Efforts have been made to address this challenge through embedding non-unitary operators into larger unitary operators in a process referred to as dilation or block encoding~\cite{Hu2020,Schlimgen2021, Schlimgen2022a, Gaikwad2022, Suri2023, Ding2024,Basile2024, Xuereb2023}, as well as using imaginary time evolution,~\cite{Kamakari2022} variational approaches,~\cite{Endo2020,Shivpuje2024, Watad2024,Luo2024,Mahdian2020a, Liu2021,Lau2023,Joo2023,Suri2018, Santos2024, Lee2021, Ollitrault2023,Gravina2024,Schlegel2023,Zhou2023} tensor train approaches,~\cite{Lyu2023} and many more.~\cite{Gupta2020b, Berreiro2011, Wei2016, Mostame2017, Garcia-Perez2020, Patsch2020, Su2020,Sun2021,Burger2022, Wang2022, Cattaneo2023, Zanetti2023, Leppaekangas2023, Guimaraes2023, Guimaraes2024,DelRe2024, Childs2017,Li2023,Li2023a,Cosacchi2018, Liu2019, Cygorek2022,Andreadakis2023,Regent2023,Muller2011,Ramusat2021, Mahdian2020b} This process is depicted schematically in Figure~\ref{fig:overview}, and with more step-by-step detail in Figure~\ref{fig:outline}.

Existing quantum algorithms have been used to explore a variety of problems relevant to chemistry,~\cite{Rost2020, Tolunay2023} biology,~\cite{Mostame2012, Gupta2020,Hu2022,Zhang2023, Oh2024, Sun2024} condensed matter,~\cite{Rost2021, Kamakari2022, Schlimgen2022b, Tornow2022,Tan2023, Liang2023} electrodynamics,~\cite{Jong2022} and materials science.~\cite{DelRe2020} The majority of these algorithms have shared a few common benchmarking systems including two-level systems with amplitude damping or dephasing channels,~\cite{Hu2020, Garcia-Perez2020, Head-Marsden2021, Schlimgen2021, Schlimgen2022a} the Jaynes-Cummings model with weak, strong, and detuned couplings,~\cite{Garcia-Perez2020, Head-Marsden2021, Warren2024} and the spin-boson model.~\cite{Wang2023} Beyond benchmarking, these algorithms have focused primarily on systems where the dynamics involve two types of processes, unitary or reversible dynamics often captured in the Hamiltonian, and non-unitary or irreversible dynamics often captured by additional terms. These initial efforts form a foundation for future applications in the treatment of open quantum systems that are classically challenging or intractable.

In this Review, we begin by providing a brief overview of the fundamental theory of open quantum systems in Section~\ref{sec:theory}. We then discuss current quantum algorithms and methodologies to consider the dynamics of OQS on quantum computers in Section~\ref{sec:methods}. Applications to fundamental open quantum system models and specific problems including photosynthetic light harvesting, the avian compass, and other systems are presented in Section~\ref{sec:apps}. Finally, we leave the reader with some concluding thoughts and an outlook towards future advances and applications in this field in Section~\ref{sec:outlook}.

\section{Theoretical Foundations}

\label{sec:theory}

In this section, we present an overview of the foundations of the theory of open quantum systems, including their definition, quantum maps, Markovian and non-Markovian dynamics, and quantum master equations.

\subsection{Defining an Open Quantum System}

\begin{figure}[h!]
    \centering
    \includegraphics[width = 0.5\textwidth]{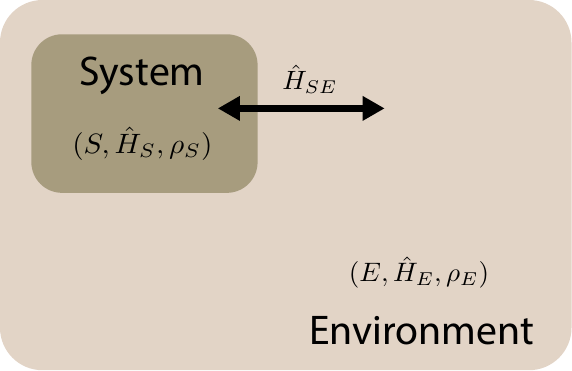}
    \caption{A quantum system partitioned into the system of interest ($S, \hat{H}_S, \rho_S$) and the environment ($E, \hat{H}_E, \rho_E$) interacting through $\hat{H}_{SE}$.}
    \label{fig:oqs}
\end{figure}

Starting from an overall closed system, we can express an open quantum system as a system $S$ embedded in an environment $E$. Here $S$ is coupled to $E$ to generate a system-environment supersystem $SE$, described for time $t$ by a density matrix $\rho_{SE}(t)$ that belongs to the Hilbert space $\mathcal{H}_{SE}=\mathcal{H}_S \otimes \mathcal{H}_E$, where dim($\mathcal{H}_E$)$\gg$dim($\mathcal{H}_S$). This is shown schematically in Figure~\ref{fig:oqs}.~\cite{Breuer2007, Rivas2012, Jagadish2019}

The time evolution of the system-environment density matrix can be expressed as a unitary operator $\hat{U}(t,t_0)$ acting on the initial density matrix $\rho_{SE}(t_0)$,
\begin{equation}
\label{eq:dyn_se}
\rho_{SE}(t) = \hat{U}(t,t_0)\rho_{SE}(t_0)\hat{U}^\dagger(t,t_0),
\end{equation}
with $t>t_0$ and $\hat{U}(t,t_0)$ defined as~\cite{Breuer2007, Rivas2012},
\begin{equation}
\label{eq:unitary}
\hat{U}(t,t_0)= {\hat T} \exp{-i\int_{t_0}^{t} \hat{H}_T(\tau) d\tau},
\end{equation}
in which ${\hat T}$ is the time-ordering operator and $\hat{H}_T$ is the total Hamiltonian of the system and the environment,
\begin{equation}
\label{eq:ht_se}
\hat{H}_T(\tau)=\hat{H}_{S}(\tau)\otimes\hat{I}_E+\hat{I}_{S}\otimes\hat{H}_E(\tau) + \hat{H}_{SE}(\tau),
\end{equation}
where $\hat{H}_{S}(\tau)$, $\hat{H}_{E}(\tau)$, and $\hat{H}_{SE}(\tau)$ are the Hamiltonians of the system, effective environment, and the interaction term between the two, respectively.~\cite{Breuer2007, Rivas2012, Jagadish2019} Formally, we can obtain the system by tracing the environment,
\begin{equation}
\label{eq:dyn_se_a}
\rho_{S}(t) = \text{Tr}_E\left(\hat{U}(t,t_0)\rho_{SE}(t_0)\hat{U}^\dagger(t,t_0)\right).
\end{equation}
Moreover, assuming that the system and the environment are not initially correlated, we can represent the embedding of the system $S$ into its environment $E$ by an extension or assignment map $E_v$~\cite{Pechukas1994,Carteret2008,Sargolzahi2020},
\begin{equation}
 E_v: \rho_{S}(t_0)\rightarrow \rho_{S}(t_0)\otimes \rho_{E}(t_0),
\end{equation}
which leads to,
\begin{align}
\label{eq:dyn_se_map}
\rho_{S}(t) = \text{Tr}_E\left(\hat{U}(t,t_0)(\rho_{S}(t_0)\otimes \rho_{E}(t_0))\hat{U}^\dagger(t,t_0)\right).
\end{align}
The above equation provides a density matrix representing our open quantum system with the only assumption being the initial separability of the system and environment.~\cite{Breuer2007,Jagadish2019,Carteret2008,Sargolzahi2020}

\subsection{Quantum Maps}

The time evolution of the system in the previous subsection defines a quantum map $\mathcal{E}_{(t,t_0)}$,
\begin{equation}
    \mathcal{E}_{(t,t_0)}: \rho_{S}(t_0)\rightarrow \rho_{S}(t), \quad (t>t_0),
\end{equation}
which is Hermitian, trace-preserving, and  positive~\cite{Breuer2007,Rivas2012,Pechukas1994,Carteret2008, Breuer2016,Jagadish2019,Manzano2020}. Unless specifically indicated in this review,  we consider quantum maps to be not only positive but also completely positive (CP).  A CP map places additional restrictions from the requirement that the system-environment density matrix remain positive semidefinite for all time~\cite{Jagadish2019, Manzano2020, Alicki1995, Pechukas1994,Pechukas1995}.

By rewriting the environment in Eq.~(\ref{eq:dyn_se_map}) as sums over the environmental states $\ket{m}$ and $\ket{n}$ with associated weights $\omega_n(t_0)$, we obtain the operator-sum representation of the system density matrix,
\begin{align}
\label{eq:quantum_maps}
\rho_{S}(t) &= \sum_{m}\bra{m}\hat{U}(t,t_0)\rho_{S}(t_0)\otimes \rho_{E}(t_0)\hat{U}^\dagger(t,t_0)\ket{m}\nonumber\\
&=\sum_{m,n}\omega_n(t_0)\bra{m}\hat{U}(t,t_0)\ket{n}\rho_{S}(t_0)\bra{n}\hat{U}^\dagger(t,t_0)\ket{m}\nonumber\\
&=\sum_{m,n}\hat{M}_{m,n}(t,t_0)\rho_{S}(t_0)\hat{M}_{m,n}^\dagger(t,t_0)\nonumber\\
&=\sum_{k}\hat{M}_{k}(t,t_0)\rho_{S}(t_0)\hat{M}_{k}^\dagger(t,t_0)
\end{align}
where $\hat{M}_{k}(t,t_0) = \hat{M}_{m,n}(t,t_0)=\sqrt{\omega_n(t_0)}\bra{m}\hat{U}(t,t_0)\ket{n}$ are the Kraus operators. These operators obey the contraction mapping,
\begin{equation}
    \sum_{k}\hat{M}_{k}(t,t_0)\hat{M}_{k}^\dagger(t,t_0)=\hat{I}_S.
\end{equation}
Importantly, this representation is not unique, as it is dependent on the basis one chooses to represent the environment~\cite{Breuer2007,Jagadish2019}. Additionally, the number of Kraus maps needed to describe fully the dynamics of the system is upper bounded by the square of the dimension of the system, dim$(\mathcal{H}_S)^2$~\cite{Jagadish2019,Suri2023}.

In general, quantum maps are not unitary. To obtain a unitary treatment, we can dilate $S$ to a larger Hilbert space according to the \textit{the Stinespring dilation theorem}~\cite{Levy2014, Ticozzi2017, Suri2023}. The theorem states that if one has a well-defined quantum map, then there exists a state $\rho_A\in\mathcal{H}_A$ where $A$ is an ancillary environment such that,
\begin{equation}
\label{eq:stinespring}
\rho_S(t) = \text{Tr}_A\left(\hat{U}_{SA}(t,t_0)\rho_{S}(t_0)\otimes \rho_{A}\hat{U}_{SA}^\dagger(t,t_0)\right),
\end{equation}
where $\hat{U}_{SA}(t,t_0)\in \mathcal{H}_S\otimes\mathcal{H}_A$ and dim($\mathcal{H}_A$) $\leq$ dim($\mathcal{H}_S$)$^2$. The unitary $\hat{U}_{SA}(t,t_0)$, is built by stacking the Kraus maps to obtain the first block column,
\begin{equation}
\hat{U}_{SA}(t,t_0)=
\begin{bmatrix}
\hat{M}_{1} & | & | &  & | \\
\vdots & \mathbf{U}_2 & \mathbf{U}_3 & \cdots & \mathbf{U}_k \\
\hat{M}_{k} & | & | &  & |  \\
\end{bmatrix},
\end{equation}
where $k\leq \text{dim}(\mathcal{H}_S)$ and the remaining block columns $\mathbf{U}_k$ are chosen such that $\hat{U}_{SA}(t,t_0)\hat{U}^\dagger _{SA}(t,t_0) = \hat{I}_{SA}$.

In addition to the Kraus maps and unitary representation, there are two other representations known as  dynamical matrices defined as~\cite{Jagadish2019, Havel2003, Kunold2024}
\begin{align}
\label{eq:A_matrix}
\mathfrak{A} &= \sum_{k}\hat{M}_{k}\otimes\hat{M}^{*}_{k}\\
\label{eq:B_matrix}
\mathfrak{B} &= \sum_{k}  \ket{\hat{M}_{k}} \bra{\hat{M}_{k}}.
\end{align}
where $\ket{{\hat{M}_{k}}}$ is the vectorized form of ${\hat{M}_{k}}$. The properties of these representations and their relationship with Kraus maps are fully described in Refs.~\citenum{Jagadish2019} and \citenum{Havel2003}. An advantage of using these representations is that they provide a direct way to determine if a map is CP; if at least one of the eigenvalues of $\mathfrak{B}$ is negative, then the quantum map is not CP.~\cite{Byrd2016}

\subsection{Quantum Master Equations}

\subsubsection{Markovian Master Equations}

A quantum map, $\mathcal{E}_{(t,t_0)}$, that describes Markovian dynamics can be represented as,
\begin{equation}
\label{eq:generator}
\mathcal{E}_{(t,t_0)}=\exp{{\mathcal{L}}(t-t_0)},
\end{equation}
where the superoperator ${\mathcal{L}}$ is the dynamics generator that acts on $\rho_S(t)$~\cite{Breuer2007, Manzano2020, Breuer2016, Alicki2007}. This leads to the time-local differential equation,
\begin{equation}
\label{eq:diff_eq}
\frac{d}{dt}\rho_S(t)={\mathcal{L}}\rho_S(t),
\end{equation}
which is referred to as the Markovian master equation.~\cite{Cubitt2012} We can derive ${\mathcal{L}}$ by projecting the Kraus maps onto an orthogonal set of operators in the Fock-Liouville space~\cite{Gorini1976, Lindblad1976, Gorini2008},
\begin{align}
\label{eq:lindblad}
\frac{d}{dt}{\rho}_S(t) &= -i\left[\hat{H},\rho_S(t)\right]
+\sum_{k=1}\gamma_k\left(\hat{L}_k\rho_S(t)\hat{L}^\dagger_k+\frac{1}{2}\{\hat{L}^\dagger_k\hat{L}_k,\rho_S(t)\}\right) \nonumber\\
&= -i\left[\hat{H},\rho_S(t)\right] + \hat{\mathcal{D}}({\rho}_S(t)),
\end{align}
where the first term is inherited from the quantum Liouville equation, and the second term is called the dissipator in which $\hat{L}_k$ are the Lindblad operators with their associated decay rates $\gamma_k$.  The notations $\{\cdot\}$  and $\left[\cdot\right]$ represent the anti-commutator and commutator, respectively. Provided the decay rates are all of the same sign, this master equation preserves the trace and complete positivity of the density matrix.

The Gorini-Kossakowski-Sudarshan-Lindblad (GKSL) equation can be derived from the Liouville equation of the combined system and environment. In this derivation, several key assumptions are at play. The system and environment are weakly coupled, known as the Born approximation, and the bath correlations decay quicker than the time scale of system relaxation, known as the Markov approximation. Often both the rotating wave and secular approximations are also invoked.~\cite{Gorini1976, Lindblad1976, Breuer2007, Manzano2020} Following a similar derivation, one can arrive at the Redfield master equation through a perturbative approximation.~\cite{Redfield1965, Breuer2007}

\subsubsection{Non-Markovian Master Equations}

The general non-Markovian master equations can be derived by projection operator techniques introduced simultaneously by Nakajima~\cite{Nakajima1958} and Zwanzig~\cite{Zwanzig1960}. The projection operator approach is based on two projection superoperators, $\mathcal{P}$ and $\mathcal{Q}$, both of which act in the composite Hilbert space $\mathcal{H}_{SE}$~\cite{Breuer2007,Haake1973,Alicki2007,Wang2023}. $\mathcal{P}$ projects $\rho_{SE}(t)$ to a state in which $S$ and $E$ are not coupled,%~\cite{Haake1973,Alicki2007,Wang2023},
\begin{equation}
\label{eq:projector_P}
\mathcal{P}\rho_{SE}(t)=\text{Tr}_E\left(\rho_{SE}(t) \right)\otimes\rho_E=\rho_S(t)\otimes\rho_E,
\end{equation}
where $\mathcal{P}\rho_{SE}(t)$ is referred to as the \textit{relevant} part of $\rho_{SE}(t)$ and $\rho_E$ represents an arbitrary state in $\mathcal{H}_E$, which is usually chosen to be a stationary state of the environment. The state resulting from $\mathcal{Q}\rho_{SE}(t)$,~\cite{Breuer2007,Alicki2007}
\begin{equation}
\label{eq:projector_Q}
\mathcal{Q}\rho_{SE}(t)=\rho_{SE}(t)-\mathcal{P}\rho_{SE}(t)
\end{equation}
is called the \textit{irrelevant} part of $\rho_{SE}(t)$. Both projection superoperators follow the same properties as one expects from a projection operator: $\mathcal{P}+\mathcal{Q}=\hat{I}_{SE}$,  $\mathcal{P}^2=\mathcal{P}$, $\mathcal{Q}^2 = \mathcal{Q}$, and $\mathcal{P}\mathcal{Q}=\mathcal{Q}\mathcal{P}=\hat{0}_{SE}$.

Using these projection operators, we can derive the generalized master equation in both non-local and local forms, often referred to as the Nakajima-Zwanzig~\cite{Nakajima1958, Zwanzig1960} equation and the time-convolutionless (TCL) equation, respectively.~\cite{Tokuyama1976, Shibata1977} The Nakajima-Zwanzig equation can be written as,\cite{Breuer2007}
\begin{align}
\label{eq:NZeq_alt_s}
\frac{d}{dt}\rho_{S}(t)&=\alpha\text{Tr}_E\left(\mathcal{P}\mathcal{L}(t)\mathcal{G}(t,t_0)\mathcal{Q}\rho_{SE}(t_0)\right) +\int^t_{t_0}ds\text{Tr}_E\left(\mathcal{K}(t,s)\mathcal{P}\rho_{SE}(s)\right),
\end{align}
where $\mathcal{G}(t,t_0)$ is a time non-unitary propagator of the form,
\begin{equation}
\label{eq:G_prop}
\mathcal{G}(t,t_0)=\hat{\textit{T}}\exp{\alpha\int^t_{t_0} ds\mathcal{Q}\mathcal{L}(s)},
\end{equation}
and $\mathcal{K}(t,s)$, which is referred to as the memory kernel, is,
\begin{equation}
\label{eq:kernel}
\mathcal{K}(t,s)=\alpha^2\mathcal{P}\mathcal{L}(t)\mathcal{G}(t,s)\mathcal{Q}\mathcal{L}(s)\mathcal{P}.
\end{equation}
Alternatively, the TCL equation can be expressed as,
\begin{align}
\label{eq:GME}
\frac{d}{dt}\rho_{S}(t)&=\hat{A}(t)\rho_S(t)+\rho_S(t)\hat{B}^\dagger(t) +\sum_i \hat{C}_i(t)\rho_S(t)\hat{D}^\dagger_i(t),
\end{align}
where $\hat{A}(t)$, $\hat{B}(t)$, $\hat{C}(t)$, and $\hat{D}(t)$ are time-dependent linear operators.

While the above approaches provide generalizations to the Markovian methods, they are challenging to solve due to computational scaling and complexity. While much progress has been made in terms of method development~\cite{Breuer2016, deVega2017}, simulating non-Markovianity efficiently and accurately remains a challenge.

\section{Algorithms for Quantum Simulation}
\label{sec:methods}

 \begin{figure}[h!]
    \centering
    \includegraphics[width = \textwidth]{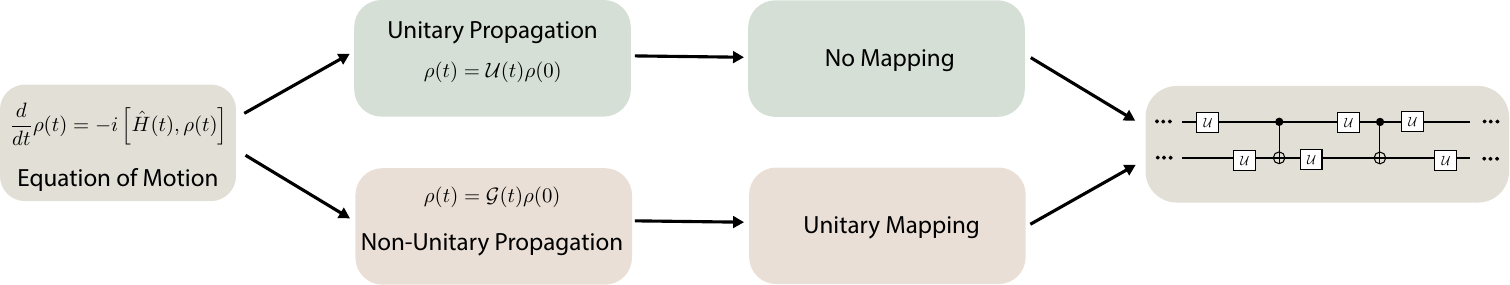}
    \caption{Simulations to open quantum systems, in comparison to closed quantum systems, introduces additional challenges such as the need for non-unitary quantum dynamics.}
    \label{fig:theory}
\end{figure}

\begin{figure}[h!]
    \centering
    \includegraphics[width = 0.5\textwidth]{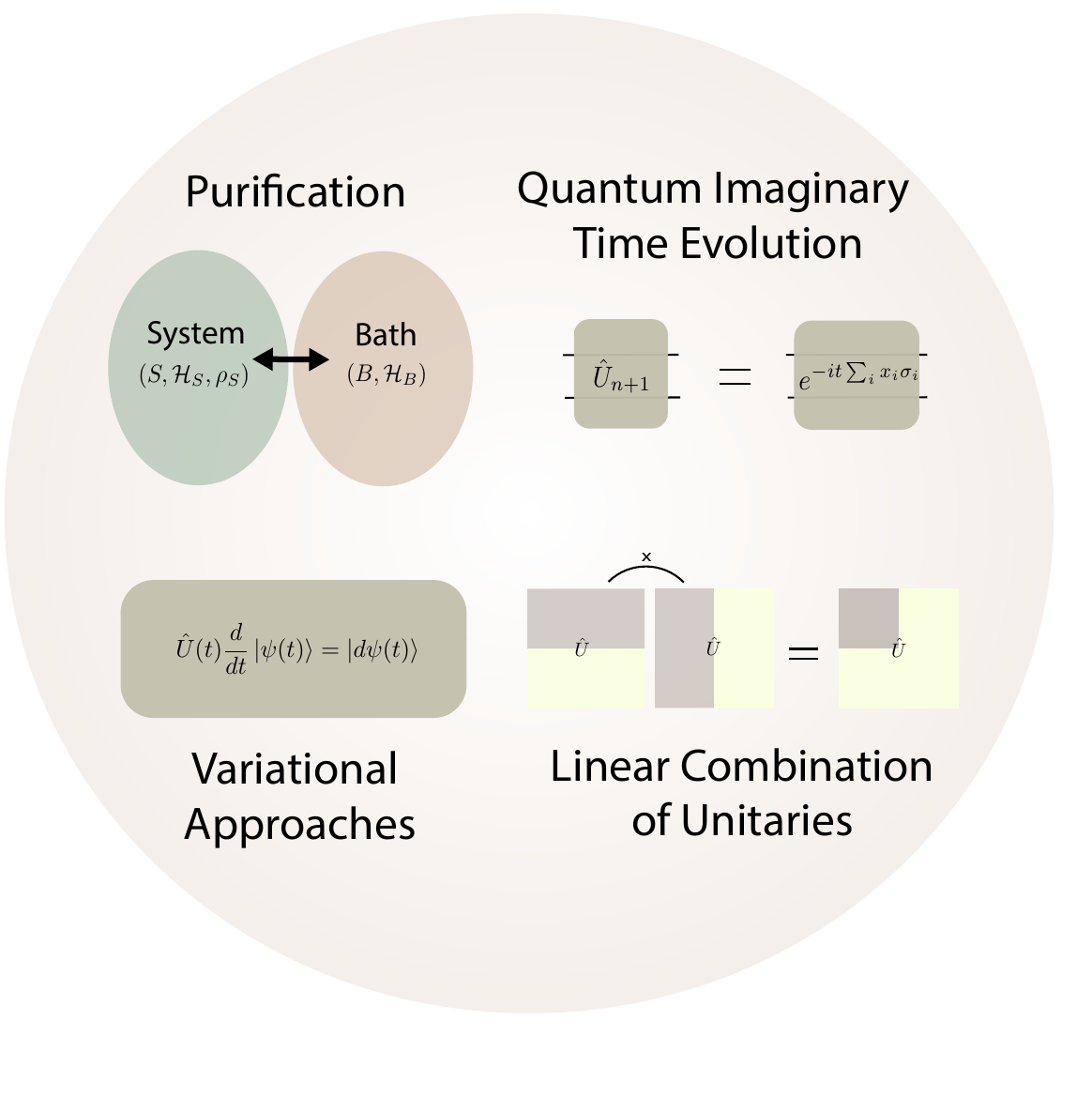}
    \caption{A selection of quantum algorithms for the simulation of open quantum systems.}
    \label{fig:methods}
\end{figure}

Quantum simulation for closed quantum systems has significant challenges, including efficient mapping of orbitals onto qubits and optimization of circuits. The extension of simulations to open quantum systems introduces additional challenges, such as the need for non-unitary quantum dynamics, as depicted schematically in Figure~\ref{fig:theory}. One of the most common approaches to this need is to translate the non-unitary dynamics into purely unitary dynamics. In this section, we review algorithms for the quantum simulation of open quantum systems, a sampling of which is depicted schematically in Figure~\ref{fig:methods}.

\subsection{Probabilistic Methods or Block-Encoding Techniques}

\begin{figure}[h!]
    \centering
    \includegraphics[width = 0.85\textwidth]{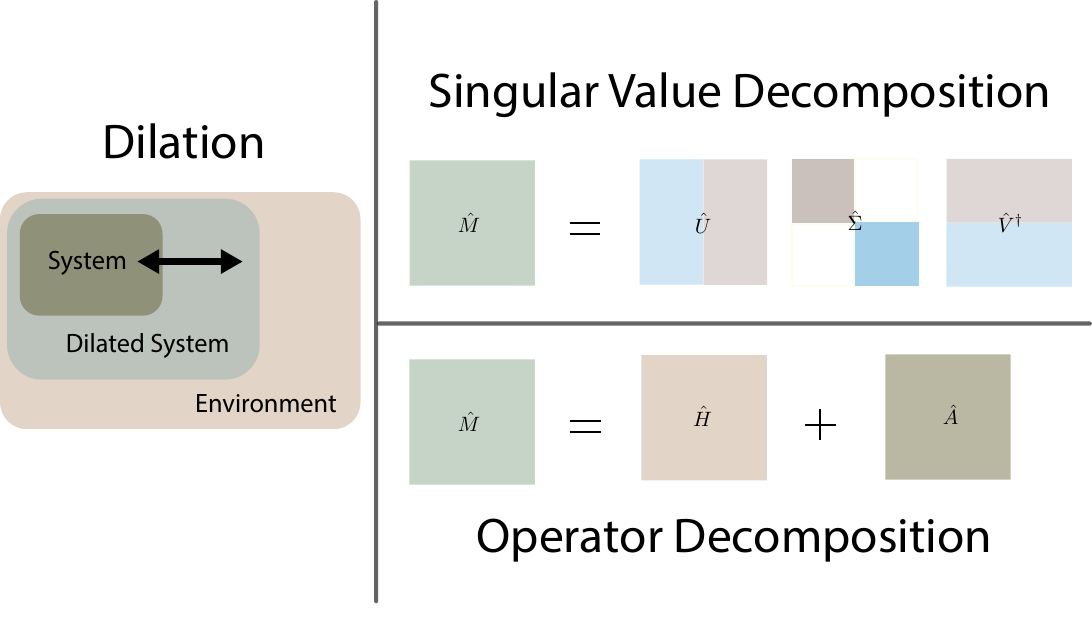}
    \caption{(Left) Block-encoding or dilation techniques where the original system Hilbert space is dilated into a larger space such that the dilated system evolves in a unitary fashion. (Right) Two examples of dilation-based approaches with the singular value decomposition approach~\cite{Schlimgen2022a} depicted schematically on the top and the operator decomposition approach~\cite{Schlimgen2021} depicted schematically on the bottom.}
    \label{fig:dilation_approaches}
\end{figure}

The general process of block encoding, or dilation, is where a non-unitary system is mapped into a unitary framework in a larger Hilbert space, as shown on the left of Figure~\ref{fig:dilation_approaches}. This unitary can be encoded into a qubit framework, where the dilation is performed with ancillary qubits. For the noisy intermediate-scale quantum (NISQ) era, in which high gate count and number of qubits can limit the usefulness of an algorithm, it is preferable to keep dilation to a minimum such that the qubit space stays as small as possible. We will discuss several different algorithms that fall within this category, including the Sz.-Nagy dilation~\cite{Hu2020,Head-Marsden2021, Hu2022, Zhang2023, Wang2023}, the unitary operator decomposition~\cite{Schlimgen2021, Schlimgen2022b}, and the classical singular value decomposition (SVD)~\cite{Schlimgen2022a, Oh2024}, of which the latter two are depicted schematically on the right of Figure~\ref{fig:dilation_approaches}.

\subsubsection{Markovian Semigroups and Linear Combination}

Some early work in understanding quantum algorithms for open quantum systems focused on finding universal sets of qubit gates for Markovian dissipative dynamics. Bacon and co-workers demonstrated that for one qubit, a single non-unitary operator parametrized by one variable is sufficient to generate all Markovian dynamics.~\cite{Bacon2001} They did so by decomposing the GKSL generator into simple components through a linear combination of semigroups and unitary conjugation. Building on this work, Sweke and co-workers described a universal set of one- and two-qubit gates to simulate arbitrary one-qubit Markovian dynamics. The resulting minimal dilation utilized techniques from Hamiltonian simulation to show complexity bounds for implementing Markovian channels with controlled accuracy. In particular, using the Suzuki-Trotter-Lie formula, the authors draw an analogy to Hamiltonian simulation to derive complexity bounds that are polynomial in the norm of the Markovian generators and the desired error tolerance.~\cite{Sweke2014, Sweke2015} This was also generalized to treat non-Markovian channels.~\cite{Sweke2016} A thorough tutorial on these approaches can be found in Ref.~\citenum{David2023}.

\subsubsection{Sz.-Nagy Dilation-Based Algorithm}

The Stinespring dilation theorem states that any non-unitary quantum operator can be converted into a unitary one~\cite{Buscemi2003, Levy2014}; however, na\"ive application of the Stinespring dilation increases the operator dimension dramatically and thus incurs high computational cost. Alternatively, the Sz.-Nagy dilation can be used to map Kraus operators into a larger unitary form that was then implementable on a quantum device~\cite{Hu2020}. This dilation works with the density matrix in operator-sum form,
\begin{equation}
    \rho(t) = \sum_k \hat{M}^{}_k\rho(0)\hat{M}_k^{\dagger},
\end{equation}
where $\rho(t)$ is the density matrix, $\hat{M}_k$ are the Kraus operators, and the summation is over all possible channels of interaction with the environment. The non-unitary operators $\hat{M}$ are dilated through the Sz.-Nagy approach such that the unitary representation is given by,
\begin{equation}
    \hat{U}_k = \begin{pmatrix}
        \hat{M}^{}_k & \hat{D}_{\hat{M}^{\dagger}_k}\\
        \hat{D}_{\hat{M}_k} & -\hat{M}^{\dagger}_k
    \end{pmatrix},
\end{equation}
where $\hat{D}_{\hat{M_k}}$ is the defect operator given by $\sqrt{I - \hat{M}^{\dagger}_k\hat{M}^{}_k}$. This dilated unitary acts on each wavefunction of a decomposed density matrix,
\begin{equation}
    \rho(t) = \sum_k\sum_j |c_j|^{2} \, \hat{U}^{}_k\lvert \psi_j \rangle \langle \psi_j \rvert \hat{U}_k^{\dagger},
\end{equation}
in which the wavefunctions are padded with zeroes to match the dimensionality of the dilated unitary operator.

We can relate the integrated Kraus representation to any differential master equation such as the Lindblad equation by defining the Kraus operators as $\hat{M}_k = \sqrt{\gamma_k\delta t}L_k$, where $L_k$ are the Lindbladian operators, with decay rates $\gamma_k$, and time step $\delta t$~\cite{Havel2003, Hu2022}. This mapping recasts the basic Sz.-Nagy dilation to apply to the Lindblad master equation. One challenge of the Sz.-Nagy dilation approach is that it incurs an undesirable classical cost in taking the matrix square root while computing the dilated unitary operator.

This approach has been effective to consider basic Kraus dynamics~\cite{Hu2020}, non-Markovian dynamics~\cite{Head-Marsden2021,Wang2023}, and the dynamics in a few biological systems~\cite{Hu2022, Zhang2023}. This algorithm has also been applied in NMR quantum processing~\cite{Gaikwad2022}.

\subsubsection{Unitary Decomposition Algorithm}

One alternative dilation approach referred to as the unitary decomposition algorithm,~\cite{Schlimgen2021} considers the decomposition of an operator into the sum of its Hermitian and anti-Hermitian components,
\begin{equation}
    \hat{M} = \hat{H} + \hat{A}.
\end{equation}
This can be recast as a sum of unitary operators
\begin{equation}
    \hat{M} = \lim_{\epsilon \rightarrow 0} \frac{1}{2\epsilon} (ie^{-i\epsilon \hat{H}} - ie^{i\epsilon \hat{H}} + e^{\epsilon \hat{A}} - e^{-\epsilon \hat{A}}).
\end{equation}
Because each of the exponentials above is a unitary operator, the equation can be implemented on a quantum computer. The summation can be performed with the linear combination of unitaries (LCU) technique~\cite{Childs2012}. While this unitary decompoition algorithm requires two ancillary qubits in the dilation in contrast to the Sz.-Nagy  dilation, it overcomes the potentially significant classical cost of computing the matrix square root of $\hat{M}$. This method has been applied using the operator-sum and the vectorized Lindblad equation~\cite{Schlimgen2021,Schlimgen2022b}.  A related dilation has been employed in electronic structure in the context of the contracted quantum eigensolvers,~\cite{Smart2024, Wang2023b, Smart2022, Smart2021} which use the residual of a contraction of the Schr{\"o}dinger equation to define an exact, iterative \emph{Ansatz} for the wave function.

\subsubsection{Classical Singular Value Decomposition Algorithm}

Another recent approach to dilating non-unitary operators efficiently relies on computing the classical singular value decomposition (SVD). This involves a classical calculation of the SVD, dilating only the singular value matrix, and implementing this new form on a quantum computer to obtain the desired quantum dynamics. The SVD is written as,
\begin{equation}
    \hat{M} = \hat{U}\hat{\Sigma}\hat{V}^\dagger,
\end{equation}
in which $\hat{U}$ and $\hat{V}^\dagger$ are unitary operators, while the $\hat{\Sigma}$ operator is non-unitary. This can be cast into unitary form via the dilation,
\begin{equation}
    \hat{U}_{\hat{\Sigma}} = \begin{pmatrix}
        \hat{\Sigma}_+ & 0 \\ 0 & \hat{\Sigma}_-
    \end{pmatrix}, %= \hat{\Sigma}_+ \oplus \hat{\Sigma}_-,
\end{equation}
where $\hat{\Sigma}_+$ and $\hat{\Sigma}_-$ are generated using the singular values of $\hat{M}$.

This approach has been applied for unnormalized state preparation, basic dynamics, and systems in quantum biology~\cite{Schlimgen2022a, Oh2024}. A recent study proposed a quantum singular value transformation algorithm to calculate the SVD on a quantum device, thus avoiding the classical complexity~\cite{Suri2023}.

\subsubsection{Monte Carlo and Minimal Dilation Approaches}

For certain classes of channels, efficient dilations can be realized, which are more amenable for quantum computers. In particular, given a quantum map of the form Eq. \eqref{eq:quantum_maps}, the channel is {unital} if it preserves the identity state, $\rho  = I_d /d$. A channel is {mixed unitary}, or {random unitary} or a {random external field}, if it has the form,
\begin{equation}
    \mathcal{E}[\rho] = \sum_i p_i U^{}_i \rho U^\dagger_i,
\end{equation}
where the collection of $p_i$ is convex and $U_i$ are unitary operators. Mixed unitary channels represent a subset of unital operators and are relevant for quantum information and quantum processes where unitary controls have some probabilistic error.

By preparing an ancilla state with probability amplitudes proportional to the square root of $p_i$, one can efficiently prepare a LCU circuit that affects the desired channel on some state~\cite{Suri2023}. However, when the number of $p_i$ is large, this can become challenging, particularly in the near-term era of quantum computation, and recent works have instead utilized Monte Carlo approaches to realize these channels.~\cite{Mazzola2024,Nagy2018,Kadowaki2018} Explicitly, an estimator of a quantum channel can be derived that samples from a multinomial distribution representing a collection of unitary channels as outcomes $\{ (p_i , U_i) \}$\cite{Peetz2023}.
The state estimator
\begin{equation}
    \hat{\rho}_\mathcal{E} = \sum_k \hat{s}_k U_k^{} \rho U_k^\dagger
\end{equation}
can be implemented with no dilation, and represents an aggregate construction of the channel by probabilistically sampling quantum circuits against a measurement outcome with a given frequency. This approach was shown to be scalable to very large systems and completely bypasses the costs associated with dilation. Additionally, successive probabilistic channels can be sampled in a Markovian fashion, allowing for time evolution of channels to be performed.

These techniques have also been applied in the context of simulating spin systems~\cite{Rost2020}. Certain non-unitary operators, such as the amplitude damping channel, can be realized as a random unitary channel on a dilated space, and thus, the above procedure can be used to avoid further dilation. Approximate random unitary channels can also be constructed for given channels.\cite{Rosgen2008}
While inherently probabilistic, the above methods have approximate success probabilities of unity, i.e., they are not post-selected and hence, scale similarly to deterministic approaches.

\subsection{Deterministic Algorithms}

In addition to block-encoding methods, a variety of deterministic methods have been developed to simulate the non-unitary evolution of quantum systems. In particular, several algorithms are based on purification techniques,~\cite{Schlimgen2022, Delgado-Granados2024} quantum imaginary time evolution (QITE),\cite{Kamakari2022} and variational approaches.~\cite{Zong2024, Shivpuje2024} Purification relies on doubling the size of the system by representing the mixed state as a wavefunction, or column vector~\cite{Kleinmann2006,Bassi2003,Hughston1993}. QITE is a measurement-based approach in which a unitary operator is generated that approximates the non-unitary evolution, typically by sampling a known decomposition of the operator.\cite{Motta2020,Kamakari2022, Endo2020, Cao2022,Mao2023,Lin2021} Finally, variational approaches utilize a parametrized circuit where the parameters are variationally optimized to simulate the time evolution.

\subsubsection{Purification}

The density-matrix purification~\cite{Schlimgen2022, Wilde2013, Nielsen2010, Inoue2018, Kleinmann2006, Bassi2003} introduces an effective bath, $B$, whose dimension equals the dimension of the system.  The openness of the system is explicitly created through the entanglement of the system with the effective bath, such that the system density matrix takes the form,
\begin{align}
\label{eq:rdm}
\rho_{S}(t) =\text{Tr}_B\left(\ket{\Psi^{SB}(t)}\bra{\Psi^{SB}(t)}\right),
\end{align}
where $\ket{\Psi^{SB}(t)}$ is a pure state of the composite system $SB$,
\begin{equation}
\label{eq:pw}
\ket{\Psi^{SB}(t)}=\sum_i \sqrt{\omega_i(t)}\ket{\Psi^S_i(t)}\ket{\Psi^B_i(t)}.
\end{equation}
This definition of the composite system allows a unitary description of its dynamics,
\begin{align}
\label{eq:upw}
\ket{\Psi^{SB}(t)} &= \hat{U}_{SB}(t,t_0)\ket{\Psi^{SB}(t_0)}\\
&={\hat T}\exp{-i\int_{t_0}^{t} \hat{H}_T(\tau) d\tau}\ket{\Psi^{SB}(t_0)},
\end{align}
where ${\hat T}$ is the time-ordering operator, and $\hat{H}_T(\tau)$ is the total Hamiltonian,
\begin{equation}
\label{eq:ht}
\hat{H}_T(\tau)=\hat{H}_{S}(\tau)\otimes\hat{I}_B+\hat{I}_{S}\otimes\hat{H}_B(\tau) + \hat{H}_{SB}(\tau).
\end{equation}
This theory results in a deterministic algorithm because the effective bath qubits are not used as postselected ancillas. Instead, the effective environment's degrees of freedom are averaged after the time propagation. Although purification doubles the number of qubits for simulation, this approach allows for treating initially entangled system-environment dynamics in Markovian and non-Markovian regimes in a common framework. It also can treat cases in which the assumption of complete positivity breaks down.~\cite{Delgado-Granados2024}

\subsubsection{Quantum Imaginary Time Evolution}

Quantum imaginary time evolution (QITE) is a general algorithm that effectively solves an eigenvalue problem by a unitary least-squares approximation of the non-unitary dynamics~\cite{Motta2020, McArdle2019}. It can also be used in open quantum systems when a propagator form of the operator is available; for example, in the vectorized Lindblad equation~\cite{Kamakari2022}. In this case, we can write the time propagation in terms of the Lindbladian $\mathcal{L}$,
\begin{equation}\label{eq:vectorized_Lindbald}
    | \rho(t) \rangle = e^{\mathcal{L}t} | \rho(0) \rangle.
\end{equation}
Here, we have used the vectorized form of the density matrix, $|\rho\rangle$. The QITE algorithm seeks a Hermitian matrix $Q$, which approximately satisfies,
\begin{equation}
    \frac{e^{\mathcal{L}t}| \rho(0)\rangle}{\|e^{\mathcal{L}t}| \rho(0)\rangle\|} = e^{-iQt} |\rho(0)\rangle.
\end{equation}
If $Q$ is written as a linear combination of Pauli operators, $\sigma_i$,
\begin{equation}
    Q = \sum_i x_i \sigma_i,
\end{equation}
then we can solve a least squares equation $Ax=b$ where,
\begin{equation}
\begin{aligned}
    A_{ij} &= \langle \rho | \sigma_i^\dagger \sigma_j |\rho \rangle, \\
    b_i &= \frac{-i}{\|e^{\mathcal{L}t}| \rho\rangle\|}\langle \rho | \sigma_i H | \rho \rangle,
\end{aligned}
\end{equation}
is measured with a quantum circuit. After finding $Q$, the vectorized time-evolved state is prepared with the unitary $e^{-iQt}$, which effectively acts as an \textit{Ansatz} for the target state.

QITE has been used to simulate open quantum systems in a variety of contexts, including dissipative two-level systems and Ising models.~\cite{Kamakari2022}  Other work has focused on reducing the circuit depth, an important development for any algorithm to be implementable on NISQ devices~\cite{Shirakawa2021, Matsushita2021, Pollmann2021}.

\subsubsection{Variational Approaches}

Time-dependent variational methods for closed or open systems rely on using a parameterized quantum circuit as an \emph{Ansatz},
\begin{eqnarray}\label{eq:ansatz_descr}
|\psi(\Vec{\theta}(t))\rangle &=& U(\Vec{\theta}(t))|0\rangle^{\otimes N},
\end{eqnarray}
where $U(\Vec{\theta}_i(t))$ corresponds to a unitary operator parameterized by a set of angles $\{ \theta_i \}$ and $N$ is the number of qubits. The variational approach is applicable to any differential equation of the following form,
\begin{eqnarray} \label{eq:generic lin_evol}
\eta(t) \frac{d |\psi(\vec{\theta}(t))\rangle}{dt} = |d \psi(\vec{\theta}(t))\rangle = \hat{\chi}(t) |\psi(\vec{\theta}(t))\rangle,
\end{eqnarray}
where $\eta(t) \in \mathbb{C}$ and $\hat{\chi}(t) \in \mathcal{L}(\mathbb{C}^{2^N})$ can be chosen to represent either closed or open system quantum dynamics.~\cite{Endo2020}

Using the traditional time-dependent variational principle, we construct the following equation of motion (EOM),~\cite{Kramer1981, Broeckhove1988, Pedersen1998, Kerman1976, Yuan2019, Hackl2020, Haegeman2013}
\begin{eqnarray} \label{eq:TDVP_evolution}
\sum_j \Im(M_{ij}(t)) \dot{\theta_j}(t) = \Im\bigg(\frac{1}{\eta(t)}V_i(t)\bigg),
\end{eqnarray}
where $M_{ij} = \langle\frac{\partial \psi(\Vec{\theta}(t))}{\partial \theta_i} | \frac{\partial \psi(\Vec{\theta}(t))}{\partial \theta_j} \rangle$, $\dot{\theta_j}(t)=\frac{d \theta_j}{dt}$, and $V_i = \langle\frac{\partial \psi(\Vec{\theta}(t))}{\partial \theta_i}| \hat{\chi}(t)|\psi(\Vec{\theta}(t))\rangle$.
For certain choices of the \emph{Ansatz}, $\Im(M(t))$ can become singular or nearly singular, leading to numerical instability~\cite{Yuan2019}. Alternative EOMs are derived using two approaches, the Dirac-Frenkel~\cite{Dirac1930, Frenkel1934, Langhoff1972} and MacLaghlan variational principles~\cite{McLachlan1964, Messina1994, Yao2021, Lee2022, Miessen2021, Alghassi2022},
\begin{eqnarray} \label{eq:D_F_evolution}
\sum_j M_{ij}(t) \dot{\theta_j}(t) = \frac{V_i(t)}{\eta(t)} \implies M(t) \dot{\Vec{\theta}}(t)= \frac{V(t)}{\eta(t)},
\end{eqnarray}
and
\begin{eqnarray} \label{eq:ML_evolution}
\sum_j \Re(M_{ij}(t)) \dot{\theta_j}(t) = \frac{V_i(t)}{\eta(t)},
\end{eqnarray}
respectively.

Whichever scheme is selected for generating the EOM for the parameters $\{\theta_j(t)\}_{j=0}^{k}$, the elements of matrix $M(t)$ and of the vector $V(t)$ can be estimated according to,
\begin{eqnarray}
 M_{ij}(t) &=&  \langle 0|^{\otimes N} \mathcal{U}_i^\dagger \mathcal{U}_j |0\rangle^{\otimes N},\label{eq:Mij_overlap_expr}
 \end{eqnarray}
 and
\begin{eqnarray}
 V_{i}(t) &=&  \langle 0|^{\otimes N} \mathcal{U}_i^\dagger \hat{\chi}(t) U(\Vec{\theta}(t)) |0\rangle^{\otimes N}. \label{eq:Vi_overlap_expr}
\end{eqnarray}
where $\mathcal{U}_i=\frac{\partial U (\Vec{\theta}(t))}{\partial \theta_i}$.  Both quantities can be estimated from the quantum circuit directly either via indirect measurements or the modified Hadamard test~\cite{Yuan2019, Endo2020, Nakaji2023}.  For the treatment of open quantum systems, the generators $\hat{\chi}(t)$ are non-Hermitian, requiring their decomposition as $\hat{\chi}_1(t) + i \hat{\chi}_2(t)$.  The generator $\hat{\chi}_1(t)$ yields a modified set of \emph{Ansatz} parameters corresponding to the nearest unitarily-evolved, parametrized state~\cite{Chen2024, Linteau2024, Shivpuje2024}.

Following the computation of $M(t)$ and $V(t)$ via quantities extracted from the quantum circuits, the EOM is solved on a classical device.~\cite{Gulliksson2000, Chen2024}  Thereafter, a new set of evolved variational parameters is suggested by the classical solver, which updates the variational \emph{Ansatz} in the quantum circuit. This process is repeated to simulate the time evolution. The variational approach requires an explicit choice of the \emph{Ansatz}, but also has the potential for compact and efficient circuits.

\section{Applications}
\label{sec:apps}

 \begin{figure}[h!]
    \centering
    \includegraphics[width = \textwidth]{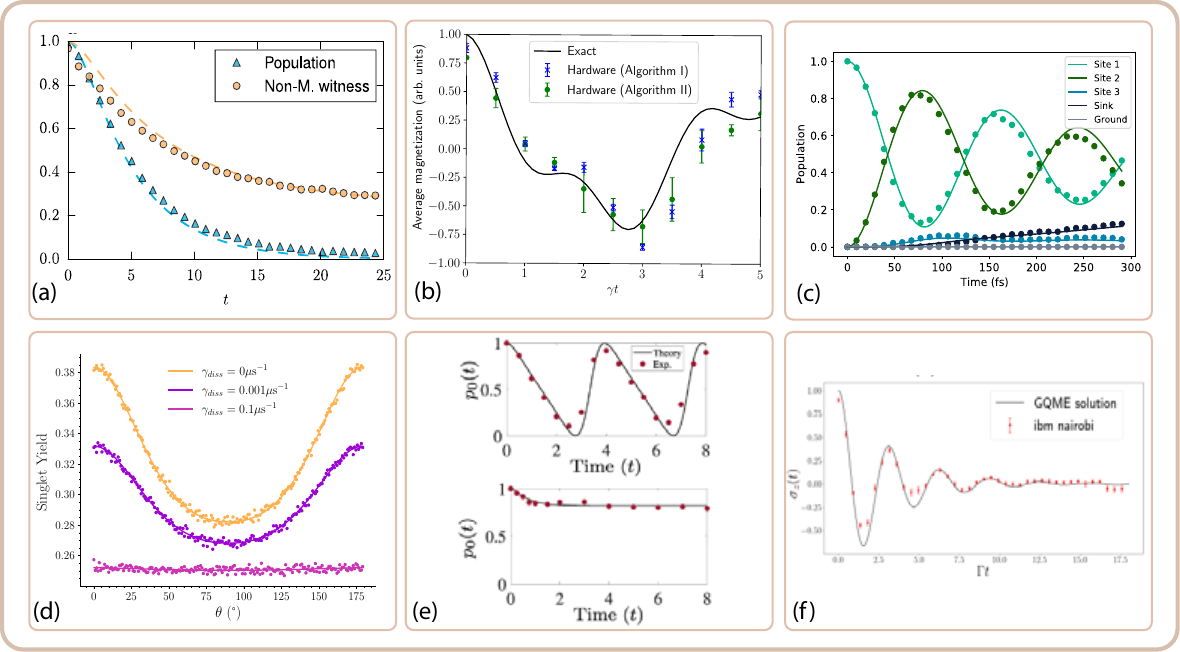}
    \caption{Quantum algorithms for the treatment of dynamics. (a) The excited-state populations (blue) and non-Markovianity witness (orange) corresponding to Markovian dynamics of a two-level system in a zero-temperature amplitude damping channel. Reprinted with permission from Ref.~\citenum{Garcia-Perez2020}. (b) Average magnetization over time of a dissipative Transverse-Field Ising Model using a quantum algorithm based on Imaginary Time Evolution. Reprinted with permission from Ref.~\citenum{Kamakari2022}. (c) Excitonic population in the Fenna-Matthews-Olson complex over time using the generalized Sz.-Nagy dilation algorithm. Reprinted with permission from Ref.~\citenum{Hu2022}. (d) The singlet yield of a radical pair mechanism of avian navigation undergoing different rates of dissipation using an SVD-based algorithm. Reprinted with permission from Ref.~\citenum{Oh2024}. (e) Ground-state populations of a parity-time symmetric system in the unbroken (top) and broken (bottom) PT-symmetric regimes using a dilation method. Reprinted with permission from Ref.~\citenum{Dogra2021}. (f) Electronic population difference between the donor and acceptor over time in the spin-boson model using a generalized Sz.-Nagy dilation algorithm. Reprinted with permission from Ref.~\citenum{Wang2023}. Copyright 2023 American Chemical Society.}
    \label{fig:applications}
\end{figure}
%\subsection{Physical Applications}

%Basile2024 simulates pauli channels -- do we want this somewhere?

In this section, we briefly review several applications of open quantum system algorithms on quantum devices, starting with a few benchmark systems. Figure~\ref{fig:applications}(a) and (b) show a two-level system in an amplitude damping channel and a two-side transverse-field Ising model, respectively.

One application is the Fenna-Matthews-Olson (FMO) complex, a widely studied biochemical complex found in green-sulfur bacteria~\cite{Fenna1975, Ishizaki2009, Zhu2012, Thyrhaug2018, Irgengioro2019,Oh2019, Kim2020, Suzuki2020, Engel2007,Lee2007}. The FMO complex is responsible for the efficient exciton energy transfer from the antenna structure to the reaction center in the photosynthetic light harvesting process~\cite{Sension2007, Barroso-Flores2017}. Studied as a model system, the detailed understanding of this trimeric-pigment protein complex can become valuable for understanding other light-harvesting complexes and for designing artificial photovoltaic systems~\cite{Hu2018a, Hu2018b}. In the first quantum simulation of the FMO dynamics, a dilation-based algorithm was used to study the FMO complex, as shown in Figure~\ref{fig:applications}(c)~\cite{Hu2022}. In the study, a Lindblad master equation with five states and seven elementary physical processes was constructed using a simplified but sufficient model provided by an earlier study showing the existence of multiple redundant pathways in the chromophore systems for transferring exciton energy to the reaction center~\cite{Skochdopole2011, Hu2022, Avdic2023,Schouten2023}. Recently, the entire seven-site system was also simulated for 2000~fs using the singular value decomposition (SVD) of the Lindblad superoperator for the system~\cite{Oh2024}. Another study examined this complex, calculating the SVD of the Kraus operators using a Walsh series representation~\cite{Seneviratne2024, Welch2014}.

To explain the unusual ability of birds to detect their orientation with respect to the Earth's magnetic field---an avian compass, the radical pair mechanism (RPM) has been proposed, which involves the reaction of a pair of radicals being correlated with each other as they are affected by magnetic fields~\cite{Ritz2004, Rodgers2009}. In the RPM, molecules in the bird’s eyes are excited by photons with certain wavelengths to generate a radical pair in the singlet state. The radical pair then converts between the singlet and triplet states as affected by both the external magnetic field and the internal nuclear spin couplings. Finally, the singlet and triplet states generate different chemical products that are sensed by the bird for directional perception. By modeling two electron spins, one nuclear spin, and two ``shelving states” representing the chemical yields of singlet and triplet states, the Lindblad master equation for the RPM of the avian compass was constructed~\cite{Gauger2011}. Both the Sz.-Nagy and SVD dilation algorithms captured the avian compass' dynamics over time, in addition to the difference in the dynamics due to varying levels of environmental noise, with sample results shown using the SVD-based algorithm in Fig.~\ref{fig:applications}(d).~\cite{Zhang2023, Oh2024}

A third application is a unique example of an open quantum system that simultaneously preserves parity and time (PT) symmetry.~\cite{Bender1998} Despite their non-Hermiticity, these systems can exhibit real eigenspectra within certain parameter regimes. In other regimes, these systems can possess complex spectra which results in intriguing topological features~\cite{ElGanainy2018,Wang2009}, such as trifold knots and braiding of the eigenstates~\cite{Patil2022}. While analog simulation has been used to investigate these systems,~\cite{Wu2019,Naghiloo2019} dilation-based algorithms have also been successful in simulating all regimes of these dynamics on quantum processors, as shown in Figure~\ref{fig:applications}(e).~\cite{Dogra2021,Kazmina2024}

The algorithms discussed in this review have been generalized to capture dynamics beyond the Markovian approximation, although these approaches are still in their infancy. The first conceptual approach relied on the Suzuki-Lie-Trotter decomposition of the system propagator, followed by a dilated implementation of the necessary unitaries~\cite{Sweke2016}. Other approaches include the use of the non-Markovian witness formulation and the Sz.-Nagy dilation applied to the Jaynes-Cummings model,~\cite{Garcia-Perez2020, Head-Marsden2021, Warren2024} the latter of which was also applied with the generalized master equation to consider the dynamics in a spin-boson model, as depicted in Figure~\ref{fig:applications}(f).~\cite{Wang2023}

\section{Conclusions and Outlook}
\label{sec:outlook}

Quantum computing holds the promise to transform our ability to solve computational problems in science and engineering that are essential to pivotal advances in society. Areas that have been targeted for quantum computing applications include optimization problems, financial modeling, artificial intelligence and machine learning, cryptography and cybersecurity, and quantum chemistry. Quantum computing in quantum chemistry may lead to markedly more efficient approaches to simulating the structure and dynamics of many-electron molecules and materials. Such improvements to the current limitations of classical computers in molecular modeling would lead to significant advances in drug discovery and design, climate modeling, energy, and material science. The vast majority of quantum computing algorithms for quantum chemistry have focused on the description and prediction of closed quantum systems, and yet most real-world systems in the applications above involve open quantum systems. This review article provides an overview of the first steps towards realizing efficient quantum computing algorithms for treating open quantum systems.

This review begins with an overview of the theory of open quantum systems in Section~\ref{sec:theory} that is applicable to both classical and quantum algorithms. The overview provides a terse but in-depth perspective on different theoretical frameworks for open quantum systems, including both Markovian and non-Markovian dynamics. Second, Section~\ref{sec:methods} discusses pioneering work in the quantum simulation of open quantum systems. Several general methodological frameworks are covered, including block-encoding techniques, density matrix purification, imaginary time evolution, and variational approaches. Third, demonstrations of these quantum algorithms in problems ranging from quantum biology to quantum materials are summarized in Section~\ref{sec:apps}, laying the groundwork for applications to a range of important problems in science and society. Despite advances made to date, this article describes only the beginning of a new frontier in the quantum simulation of open quantum systems. We hope that it will serve as a resource and inspiration for future advances that will take us closer to realizing the potential of quantum computing in chemistry.

\section{Acknowledgments}

D.A.M., P.N., and S.K. gratefully acknowledge funding from the Department of Energy, Office of Basic Energy Sciences, Grant No. DE-SC0019215.  D.A.M. and P.N. acknowledge the U.S. National Science Foundation (NSF) Grant No. DMR-2037783, and D.A.M. acknowledges the U.S. NSF Grant No. CHE-2155082 and the NSF QuBBE Quantum Leap Challenge Institute (NSF OMA-2121044). S.K. acknowledges the financial support of the National Science Foundation under award number 2124511, CCI Phase I: NSF Center for Quantum Dynamics on Modular Quantum Devices (CQD-MQD) and financial support from the Quantum Science Center, a quantum partnership funded by the U.S. Department of Energy (DOE).  K.H.M. acknowledges start-up funding from Washington University in St. Louis and the University of Minnesota.  L.H.D.G. acknowledges that this material is based upon work supported by the U.S. Department of Energy, Office of Science, Office of Advanced Scientific Computing Research, Department of Energy Computational Science Graduate Fellowship under Award Number DE-SC0024386. I.A. acknowledges support from the NSF Graduate Research Fellowship Program under Grant No. 2140001.

\bibliography{main}

\providecommand{\latin}[1]{#1}
\makeatletter
\providecommand{\doi}
  {\begingroup\let\do\@makeother\dospecials
  \catcode`\{=1 \catcode`\}=2 \doi@aux}
\providecommand{\doi@aux}[1]{\endgroup\texttt{#1}}
\makeatother
\providecommand*\mcitethebibliography{\thebibliography}
\csname @ifundefined\endcsname{endmcitethebibliography}
  {\let\endmcitethebibliography\endthebibliography}{}
\begin{mcitethebibliography}{213}
\providecommand*\natexlab[1]{#1}
\providecommand*\mciteSetBstSublistMode[1]{}
\providecommand*\mciteSetBstMaxWidthForm[2]{}
\providecommand*\mciteBstWouldAddEndPuncttrue
  {\def\EndOfBibitem{\unskip.}}
\providecommand*\mciteBstWouldAddEndPunctfalse
  {\let\EndOfBibitem\relax}
\providecommand*\mciteSetBstMidEndSepPunct[3]{}
\providecommand*\mciteSetBstSublistLabelBeginEnd[3]{}
\providecommand*\EndOfBibitem{}
\mciteSetBstSublistMode{f}
\mciteSetBstMaxWidthForm{subitem}{(\alph{mcitesubitemcount})}
\mciteSetBstSublistLabelBeginEnd
  {\mcitemaxwidthsubitemform\space}
  {\relax}
  {\relax}

\bibitem[Breuer and Petruccione(2007)Breuer, and Petruccione]{Breuer2007}
Breuer,~H.-P.; Petruccione,~F. \emph{The Theory of Open Quantum Systems};
  Oxford University Press, 2007\relax
\mciteBstWouldAddEndPuncttrue
\mciteSetBstMidEndSepPunct{\mcitedefaultmidpunct}
{\mcitedefaultendpunct}{\mcitedefaultseppunct}\relax
\EndOfBibitem
\bibitem[Rivas and Huelga(2012)Rivas, and Huelga]{Rivas2012}
Rivas,~A.; Huelga,~S.~F. \emph{Open quantum systems}; Springer, 2012;
  Vol.~10\relax
\mciteBstWouldAddEndPuncttrue
\mciteSetBstMidEndSepPunct{\mcitedefaultmidpunct}
{\mcitedefaultendpunct}{\mcitedefaultseppunct}\relax
\EndOfBibitem
\bibitem[Breuer \latin{et~al.}(2016)Breuer, Laine, Piilo, and
  Vacchini]{Breuer2016}
Breuer,~H.-P.; Laine,~E.-M.; Piilo,~J.; Vacchini,~B. Colloquium: Non-Markovian
  dynamics in open quantum systems. \emph{Rev. Mod. Phys.} \textbf{2016},
  \emph{88}, 021002\relax
\mciteBstWouldAddEndPuncttrue
\mciteSetBstMidEndSepPunct{\mcitedefaultmidpunct}
{\mcitedefaultendpunct}{\mcitedefaultseppunct}\relax
\EndOfBibitem
\bibitem[Ángel Rivas \latin{et~al.}(2014)Ángel Rivas, Huelga, and
  Plenio]{Rivas2014}
Ángel Rivas,; Huelga,~S.~F.; Plenio,~M.~B. Quantum non-Markovianity:
  characterization, quantification and detection. \emph{Reports on Progress in
  Physics} \textbf{2014}, \emph{77}, 094001\relax
\mciteBstWouldAddEndPuncttrue
\mciteSetBstMidEndSepPunct{\mcitedefaultmidpunct}
{\mcitedefaultendpunct}{\mcitedefaultseppunct}\relax
\EndOfBibitem
\bibitem[Li \latin{et~al.}(2019)Li, Guo, and Piilo]{Li2019a}
Li,~C.-F.; Guo,~G.-C.; Piilo,~J. Non-Markovian quantum dynamics: What does it
  mean? \emph{Europhysics Letters} \textbf{2019}, \emph{127}, 50001\relax
\mciteBstWouldAddEndPuncttrue
\mciteSetBstMidEndSepPunct{\mcitedefaultmidpunct}
{\mcitedefaultendpunct}{\mcitedefaultseppunct}\relax
\EndOfBibitem
\bibitem[Li \latin{et~al.}(2020)Li, Guo, and Piilo]{Li2019b}
Li,~C.-F.; Guo,~G.-C.; Piilo,~J. Non-Markovian quantum dynamics: What is it
  good for? \emph{Europhysics Letters} \textbf{2020}, \emph{128}, 30001\relax
\mciteBstWouldAddEndPuncttrue
\mciteSetBstMidEndSepPunct{\mcitedefaultmidpunct}
{\mcitedefaultendpunct}{\mcitedefaultseppunct}\relax
\EndOfBibitem
\bibitem[Kraus \latin{et~al.}(2008)Kraus, B\"uchler, Diehl, Kantian, Micheli,
  and Zoller]{Kraus2008}
Kraus,~B.; B\"uchler,~H.~P.; Diehl,~S.; Kantian,~A.; Micheli,~A.; Zoller,~P.
  Preparation of entangled states by quantum Markov processes. \emph{Physical
  Review A} \textbf{2008}, \emph{78}, 042307\relax
\mciteBstWouldAddEndPuncttrue
\mciteSetBstMidEndSepPunct{\mcitedefaultmidpunct}
{\mcitedefaultendpunct}{\mcitedefaultseppunct}\relax
\EndOfBibitem
\bibitem[Diehl \latin{et~al.}(2008)Diehl, Micheli, Kantian, Kraus, B{\"u}chler,
  and Zoller]{Diehl2008}
Diehl,~S.; Micheli,~A.; Kantian,~A.; Kraus,~B.; B{\"u}chler,~H.~P.; Zoller,~P.
  Quantum states and phases in driven open quantum systems with cold atoms.
  \emph{Nature Physics} \textbf{2008}, \emph{4}, 878--883\relax
\mciteBstWouldAddEndPuncttrue
\mciteSetBstMidEndSepPunct{\mcitedefaultmidpunct}
{\mcitedefaultendpunct}{\mcitedefaultseppunct}\relax
\EndOfBibitem
\bibitem[Verstraete \latin{et~al.}(2009)Verstraete, Wolf, and
  Ignacio~Cirac]{Verstraete2009}
Verstraete,~F.; Wolf,~M.~M.; Ignacio~Cirac,~J. Quantum computation and
  quantum-state engineering driven by dissipation. \emph{Nature Physics}
  \textbf{2009}, \emph{5}, 633--636\relax
\mciteBstWouldAddEndPuncttrue
\mciteSetBstMidEndSepPunct{\mcitedefaultmidpunct}
{\mcitedefaultendpunct}{\mcitedefaultseppunct}\relax
\EndOfBibitem
\bibitem[Head-Marsden \latin{et~al.}(2021)Head-Marsden, Flick, Ciccarino, and
  Narang]{HeadMarsdenFlick2021}
Head-Marsden,~K.; Flick,~J.; Ciccarino,~C.~J.; Narang,~P. Quantum information
  and algorithms for correlated quantum matter. \emph{Chemical Reviews}
  \textbf{2021}, \emph{121}, 3061--3120\relax
\mciteBstWouldAddEndPuncttrue
\mciteSetBstMidEndSepPunct{\mcitedefaultmidpunct}
{\mcitedefaultendpunct}{\mcitedefaultseppunct}\relax
\EndOfBibitem
\bibitem[Olivera-Atencio \latin{et~al.}(2023)Olivera-Atencio, Lamata, and
  Casado-Pascual]{Olivera-Atencio2023}
Olivera-Atencio,~M.~L.; Lamata,~L.; Casado-Pascual,~J. Benefits of open quantum
  systems for quantum machine learning. \emph{Advanced Quantum Technologies}
  \textbf{2023}, \emph{n/a}, 2300247\relax
\mciteBstWouldAddEndPuncttrue
\mciteSetBstMidEndSepPunct{\mcitedefaultmidpunct}
{\mcitedefaultendpunct}{\mcitedefaultseppunct}\relax
\EndOfBibitem
\bibitem[Gisin \latin{et~al.}(2002)Gisin, Ribordy, Tittel, and
  Zbinden]{Gisin2002}
Gisin,~N.; Ribordy,~G.; Tittel,~W.; Zbinden,~H. Quantum cryptography.
  \emph{Reviews of Modern Physics} \textbf{2002}, \emph{74}, 145–195\relax
\mciteBstWouldAddEndPuncttrue
\mciteSetBstMidEndSepPunct{\mcitedefaultmidpunct}
{\mcitedefaultendpunct}{\mcitedefaultseppunct}\relax
\EndOfBibitem
\bibitem[Kolkowitz \latin{et~al.}(2016)Kolkowitz, Bromley, Bothwell, Wall,
  Marti, Koller, Zhang, Rey, and Ye]{Kolkowitz2016}
Kolkowitz,~S.; Bromley,~S.~L.; Bothwell,~T.; Wall,~M.~L.; Marti,~G.~E.;
  Koller,~A.~P.; Zhang,~X.; Rey,~A.~M.; Ye,~J. Spin–orbit-coupled fermions in
  an optical lattice clock. \emph{Nature} \textbf{2016}, \emph{542},
  66–70\relax
\mciteBstWouldAddEndPuncttrue
\mciteSetBstMidEndSepPunct{\mcitedefaultmidpunct}
{\mcitedefaultendpunct}{\mcitedefaultseppunct}\relax
\EndOfBibitem
\bibitem[Martin \latin{et~al.}(2013)Martin, Bishof, Swallows, Zhang, Benko, von
  Stecher, Gorshkov, Rey, and Ye]{Martin2013}
Martin,~M.~J.; Bishof,~M.; Swallows,~M.~D.; Zhang,~X.; Benko,~C.; von
  Stecher,~J.; Gorshkov,~A.~V.; Rey,~A.~M.; Ye,~J. A Quantum Many-Body Spin
  System in an Optical Lattice Clock. \emph{Science} \textbf{2013}, \emph{341},
  632–636\relax
\mciteBstWouldAddEndPuncttrue
\mciteSetBstMidEndSepPunct{\mcitedefaultmidpunct}
{\mcitedefaultendpunct}{\mcitedefaultseppunct}\relax
\EndOfBibitem
\bibitem[Georgescu \latin{et~al.}(2014)Georgescu, Ashhab, and
  Nori]{Georgescu2014}
Georgescu,~I.; Ashhab,~S.; Nori,~F. Quantum simulation. \emph{Reviews of Modern
  Physics} \textbf{2014}, \emph{86}, 153–185\relax
\mciteBstWouldAddEndPuncttrue
\mciteSetBstMidEndSepPunct{\mcitedefaultmidpunct}
{\mcitedefaultendpunct}{\mcitedefaultseppunct}\relax
\EndOfBibitem
\bibitem[Yunger~Halpern \latin{et~al.}(2016)Yunger~Halpern, Faist, Oppenheim,
  and Winter]{YungerHalpern2016}
Yunger~Halpern,~N.; Faist,~P.; Oppenheim,~J.; Winter,~A. Microcanonical and
  resource-theoretic derivations of the thermal state of a quantum system with
  noncommuting charges. \emph{Nature Communications} \textbf{2016},
  \emph{7}\relax
\mciteBstWouldAddEndPuncttrue
\mciteSetBstMidEndSepPunct{\mcitedefaultmidpunct}
{\mcitedefaultendpunct}{\mcitedefaultseppunct}\relax
\EndOfBibitem
\bibitem[Wiseman and Milburn(2009)Wiseman, and Milburn]{Wiseman2009}
Wiseman,~H.~M.; Milburn,~G.~J. \emph{Quantum Measurement and Control};
  Cambridge University Press, 2009\relax
\mciteBstWouldAddEndPuncttrue
\mciteSetBstMidEndSepPunct{\mcitedefaultmidpunct}
{\mcitedefaultendpunct}{\mcitedefaultseppunct}\relax
\EndOfBibitem
\bibitem[Pechen and Rabitz(2014)Pechen, and Rabitz]{Pechen2014}
Pechen,~A.; Rabitz,~H. {Incoherent Control of Open Quantum Systems}.
  \emph{Journal of Mathematical Sciences} \textbf{2014}, \emph{199},
  695--701\relax
\mciteBstWouldAddEndPuncttrue
\mciteSetBstMidEndSepPunct{\mcitedefaultmidpunct}
{\mcitedefaultendpunct}{\mcitedefaultseppunct}\relax
\EndOfBibitem
\bibitem[Wu \latin{et~al.}(2007)Wu, Pechen, Brif, and Rabitz]{Wu2007}
Wu,~R.; Pechen,~A.; Brif,~C.; Rabitz,~H. {Controllability of open quantum
  systems with Kraus-map dynamics}. \emph{Journal of Physics A: Mathematical
  and Theoretical} \textbf{2007}, \emph{40}, 5681--5693\relax
\mciteBstWouldAddEndPuncttrue
\mciteSetBstMidEndSepPunct{\mcitedefaultmidpunct}
{\mcitedefaultendpunct}{\mcitedefaultseppunct}\relax
\EndOfBibitem
\bibitem[Avdic \latin{et~al.}(2023)Avdic, Sager-Smith, and
  Mazziotti]{Avdic2023b}
Avdic,~I.; Sager-Smith,~L.~M.; Mazziotti,~D.~A. {Open quantum system violates
  generalized Pauli constraints on quantum device}. \emph{Communications
  Physics} \textbf{2023}, \emph{6}, 180\relax
\mciteBstWouldAddEndPuncttrue
\mciteSetBstMidEndSepPunct{\mcitedefaultmidpunct}
{\mcitedefaultendpunct}{\mcitedefaultseppunct}\relax
\EndOfBibitem
\bibitem[Mohseni \latin{et~al.}(2014)Mohseni, Omar, Engel, and
  Plenio]{Mohseni2014}
Mohseni,~M.; Omar,~Y.; Engel,~G.~S.; Plenio,~M.~B. \emph{Quantum effects in
  biology}; Cambridge University Press, 2014\relax
\mciteBstWouldAddEndPuncttrue
\mciteSetBstMidEndSepPunct{\mcitedefaultmidpunct}
{\mcitedefaultendpunct}{\mcitedefaultseppunct}\relax
\EndOfBibitem
\bibitem[Mohseni \latin{et~al.}(2013)Mohseni, Shabani, Lloyd, Omar, and
  Rabitz]{Mohseni2013}
Mohseni,~M.; Shabani,~A.; Lloyd,~S.; Omar,~Y.; Rabitz,~H. {Geometrical effects
  on energy transfer in disordered open quantum systems}. \emph{The Journal of
  Chemical Physics} \textbf{2013}, \emph{138}, 204309\relax
\mciteBstWouldAddEndPuncttrue
\mciteSetBstMidEndSepPunct{\mcitedefaultmidpunct}
{\mcitedefaultendpunct}{\mcitedefaultseppunct}\relax
\EndOfBibitem
\bibitem[Kitai \latin{et~al.}(2020)Kitai, Guo, Ju, Tanaka, Tsuda, Shiomi, and
  Tamura]{Kitai2020}
Kitai,~K.; Guo,~J.; Ju,~S.; Tanaka,~S.; Tsuda,~K.; Shiomi,~J.; Tamura,~R.
  Designing metamaterials with quantum annealing and factorization machines.
  \emph{Physical Review Research} \textbf{2020}, \emph{2}\relax
\mciteBstWouldAddEndPuncttrue
\mciteSetBstMidEndSepPunct{\mcitedefaultmidpunct}
{\mcitedefaultendpunct}{\mcitedefaultseppunct}\relax
\EndOfBibitem
\bibitem[Nielsen and Chuang(2010)Nielsen, and Chuang]{Nielsen2010}
Nielsen,~M.~A.; Chuang,~I.~L. \emph{Quantum computation and quantum
  information}; Cambridge university press, 2010\relax
\mciteBstWouldAddEndPuncttrue
\mciteSetBstMidEndSepPunct{\mcitedefaultmidpunct}
{\mcitedefaultendpunct}{\mcitedefaultseppunct}\relax
\EndOfBibitem
\bibitem[Wilde(2013)]{Wilde2013}
Wilde,~M.~M. \emph{Quantum information theory}; Cambridge university press,
  2013\relax
\mciteBstWouldAddEndPuncttrue
\mciteSetBstMidEndSepPunct{\mcitedefaultmidpunct}
{\mcitedefaultendpunct}{\mcitedefaultseppunct}\relax
\EndOfBibitem
\bibitem[Kassal \latin{et~al.}(2011)Kassal, Whitfield, Perdomo-Ortiz, Yung, and
  Aspuru-Guzik]{Kassal2011}
Kassal,~I.; Whitfield,~J.~D.; Perdomo-Ortiz,~A.; Yung,~M.-H.; Aspuru-Guzik,~A.
  Simulating chemistry using quantum computers. \emph{Annual review of physical
  chemistry} \textbf{2011}, \emph{62}, 185--207\relax
\mciteBstWouldAddEndPuncttrue
\mciteSetBstMidEndSepPunct{\mcitedefaultmidpunct}
{\mcitedefaultendpunct}{\mcitedefaultseppunct}\relax
\EndOfBibitem
\bibitem[Cao(2019)]{Cao2019}
Cao,~Y.,~et~al. Quantum Chemistry in the Age of Quantum Computing.
  \emph{Chemical Reviews} \textbf{2019}, \emph{119}, 10856--10915\relax
\mciteBstWouldAddEndPuncttrue
\mciteSetBstMidEndSepPunct{\mcitedefaultmidpunct}
{\mcitedefaultendpunct}{\mcitedefaultseppunct}\relax
\EndOfBibitem
\bibitem[Tilly \latin{et~al.}(2022)Tilly, Chen, Cao, Picozzi, Setia, Li, Grant,
  Wossnig, Rungger, Booth, and Tennyson]{Tilly2022}
Tilly,~J.; Chen,~H.; Cao,~S.; Picozzi,~D.; Setia,~K.; Li,~Y.; Grant,~E.;
  Wossnig,~L.; Rungger,~I.; Booth,~G.~H.; Tennyson,~J. The Variational Quantum
  Eigensolver: A review of methods and best practices. \emph{Physics Reports}
  \textbf{2022}, \emph{986}, 1--128\relax
\mciteBstWouldAddEndPuncttrue
\mciteSetBstMidEndSepPunct{\mcitedefaultmidpunct}
{\mcitedefaultendpunct}{\mcitedefaultseppunct}\relax
\EndOfBibitem
\bibitem[Miessen \latin{et~al.}(2023)Miessen, Ollitrault, Tacchino, and
  Tavernelli]{Miessen2023}
Miessen,~A.; Ollitrault,~P.~J.; Tacchino,~F.; Tavernelli,~I. Quantum algorithms
  for quantum dynamics. \emph{Nature Computational Science} \textbf{2023},
  \emph{3}, 25--37\relax
\mciteBstWouldAddEndPuncttrue
\mciteSetBstMidEndSepPunct{\mcitedefaultmidpunct}
{\mcitedefaultendpunct}{\mcitedefaultseppunct}\relax
\EndOfBibitem
\bibitem[Fauseweh(2024)]{Fauseweh2024}
Fauseweh,~B. Quantum many-body simulations on digital quantum computers:
  State-of-the-art and future challenges. \emph{Nature Communications}
  \textbf{2024}, \emph{15}\relax
\mciteBstWouldAddEndPuncttrue
\mciteSetBstMidEndSepPunct{\mcitedefaultmidpunct}
{\mcitedefaultendpunct}{\mcitedefaultseppunct}\relax
\EndOfBibitem
\bibitem[Cerezo \latin{et~al.}(2021)Cerezo, Arrasmith, Babbush, Benjamin, Endo,
  Fujii, McClean, Mitarai, Yuan, Cincio, \latin{et~al.} others]{Cerezo2021}
Cerezo,~M.; Arrasmith,~A.; Babbush,~R.; Benjamin,~S.~C.; Endo,~S.; Fujii,~K.;
  McClean,~J.~R.; Mitarai,~K.; Yuan,~X.; Cincio,~L., \latin{et~al.}
  Variational quantum algorithms. \emph{Nature Reviews Physics} \textbf{2021},
  \emph{3}, 625--644\relax
\mciteBstWouldAddEndPuncttrue
\mciteSetBstMidEndSepPunct{\mcitedefaultmidpunct}
{\mcitedefaultendpunct}{\mcitedefaultseppunct}\relax
\EndOfBibitem
\bibitem[Bharti \latin{et~al.}(2022)Bharti, Cervera-Lierta, Kyaw, Haug,
  Alperin-Lea, Anand, Degroote, Heimonen, Kottmann, Menke, \latin{et~al.}
  others]{Bharti2022}
Bharti,~K.; Cervera-Lierta,~A.; Kyaw,~T.~H.; Haug,~T.; Alperin-Lea,~S.;
  Anand,~A.; Degroote,~M.; Heimonen,~H.; Kottmann,~J.~S.; Menke,~T.,
  \latin{et~al.}  Noisy intermediate-scale quantum algorithms. \emph{Reviews of
  Modern Physics} \textbf{2022}, \emph{94}, 015004\relax
\mciteBstWouldAddEndPuncttrue
\mciteSetBstMidEndSepPunct{\mcitedefaultmidpunct}
{\mcitedefaultendpunct}{\mcitedefaultseppunct}\relax
\EndOfBibitem
\bibitem[Wecker \latin{et~al.}(2015)Wecker, Hastings, and Troyer]{Wecker2015}
Wecker,~D.; Hastings,~M.~B.; Troyer,~M. Progress towards practical quantum
  variational algorithms. \emph{Phys. Rev. A} \textbf{2015}, \emph{92},
  042303\relax
\mciteBstWouldAddEndPuncttrue
\mciteSetBstMidEndSepPunct{\mcitedefaultmidpunct}
{\mcitedefaultendpunct}{\mcitedefaultseppunct}\relax
\EndOfBibitem
\bibitem[Huang \latin{et~al.}(2020)Huang, Wu, Fan, and Zhu]{Huang2020}
Huang,~H.-L.; Wu,~D.; Fan,~D.; Zhu,~X. Superconducting quantum computing: a
  review. \emph{Science China Information Sciences} \textbf{2020},
  \emph{63}\relax
\mciteBstWouldAddEndPuncttrue
\mciteSetBstMidEndSepPunct{\mcitedefaultmidpunct}
{\mcitedefaultendpunct}{\mcitedefaultseppunct}\relax
\EndOfBibitem
\bibitem[Bruzewicz \latin{et~al.}(2019)Bruzewicz, Chiaverini, McConnell, and
  Sage]{Bruzewicz2019}
Bruzewicz,~C.~D.; Chiaverini,~J.; McConnell,~R.; Sage,~J.~M. {Trapped-ion
  quantum computing: Progress and challenges}. \emph{Applied Physics Reviews}
  \textbf{2019}, \emph{6}, 021314\relax
\mciteBstWouldAddEndPuncttrue
\mciteSetBstMidEndSepPunct{\mcitedefaultmidpunct}
{\mcitedefaultendpunct}{\mcitedefaultseppunct}\relax
\EndOfBibitem
\bibitem[Wintersperger \latin{et~al.}(2023)Wintersperger, Dommert, Ehmer,
  Hoursanov, Klepsch, Mauerer, Reuber, Strohm, Yin, and
  Luber]{Wintersperger2023}
Wintersperger,~K.; Dommert,~F.; Ehmer,~T.; Hoursanov,~A.; Klepsch,~J.;
  Mauerer,~W.; Reuber,~G.; Strohm,~T.; Yin,~M.; Luber,~S. Neutral atom quantum
  computing hardware: performance and end-user perspective. \emph{EPJ Quantum
  Technology} \textbf{2023}, \emph{10}, 32\relax
\mciteBstWouldAddEndPuncttrue
\mciteSetBstMidEndSepPunct{\mcitedefaultmidpunct}
{\mcitedefaultendpunct}{\mcitedefaultseppunct}\relax
\EndOfBibitem
\bibitem[Bacon \latin{et~al.}(2001)Bacon, Childs, Chuang, Kempe, Leung, and
  Zhou]{Bacon2001}
Bacon,~D.; Childs,~A.~M.; Chuang,~I.~L.; Kempe,~J.; Leung,~D.~W.; Zhou,~X.
  Universal simulation of Markovian quantum dynamics. \emph{Phys. Rev. A}
  \textbf{2001}, \emph{64}, 062302\relax
\mciteBstWouldAddEndPuncttrue
\mciteSetBstMidEndSepPunct{\mcitedefaultmidpunct}
{\mcitedefaultendpunct}{\mcitedefaultseppunct}\relax
\EndOfBibitem
\bibitem[Wang \latin{et~al.}(2011)Wang, Ashhab, and Nori]{Wang2011}
Wang,~H.; Ashhab,~S.; Nori,~F. Quantum algorithm for simulating the dynamics of
  an open quantum system. \emph{Physical Review A} \textbf{2011},
  \emph{83}\relax
\mciteBstWouldAddEndPuncttrue
\mciteSetBstMidEndSepPunct{\mcitedefaultmidpunct}
{\mcitedefaultendpunct}{\mcitedefaultseppunct}\relax
\EndOfBibitem
\bibitem[Sweke \latin{et~al.}(2015)Sweke, Sinayskiy, Bernard, and
  Petruccione]{Sweke2015}
Sweke,~R.; Sinayskiy,~I.; Bernard,~D.; Petruccione,~F. Universal simulation of
  Markovian open quantum systems. \emph{Phys. Rev. A} \textbf{2015}, \emph{91},
  062308\relax
\mciteBstWouldAddEndPuncttrue
\mciteSetBstMidEndSepPunct{\mcitedefaultmidpunct}
{\mcitedefaultendpunct}{\mcitedefaultseppunct}\relax
\EndOfBibitem
\bibitem[Sweke \latin{et~al.}(2016)Sweke, Sanz, Sinayskiy, Petruccione, and
  Solano]{Sweke2016}
Sweke,~R.; Sanz,~M.; Sinayskiy,~I.; Petruccione,~F.; Solano,~E. Digital quantum
  simulation of many-body non-Markovian dynamics. \emph{Phys. Rev. A}
  \textbf{2016}, \emph{94}, 022317\relax
\mciteBstWouldAddEndPuncttrue
\mciteSetBstMidEndSepPunct{\mcitedefaultmidpunct}
{\mcitedefaultendpunct}{\mcitedefaultseppunct}\relax
\EndOfBibitem
\bibitem[Wei \latin{et~al.}(2016)Wei, Ruan, and Long]{Wei2016}
Wei,~S.-J.; Ruan,~D.; Long,~G.-L. Duality quantum algorithm efficiently
  simulates open quantum systems. \emph{Scientific Reports} \textbf{2016},
  \emph{6}, 30727\relax
\mciteBstWouldAddEndPuncttrue
\mciteSetBstMidEndSepPunct{\mcitedefaultmidpunct}
{\mcitedefaultendpunct}{\mcitedefaultseppunct}\relax
\EndOfBibitem
\bibitem[Hu \latin{et~al.}(2020)Hu, Xia, and Kais]{Hu2020}
Hu,~Z.; Xia,~R.; Kais,~S. A quantum algorithm for evolving open quantum
  dynamics on quantum computing devices. \emph{Sci. Rep.} \textbf{2020},
  \emph{10}, 3301\relax
\mciteBstWouldAddEndPuncttrue
\mciteSetBstMidEndSepPunct{\mcitedefaultmidpunct}
{\mcitedefaultendpunct}{\mcitedefaultseppunct}\relax
\EndOfBibitem
\bibitem[Patsch \latin{et~al.}(2020)Patsch, Maniscalco, and Koch]{Patsch2020}
Patsch,~S.; Maniscalco,~S.; Koch,~C.~P. Simulation of open-quantum-system
  dynamics using the quantum Zeno effect. \emph{Physical Review Research}
  \textbf{2020}, \emph{2}, 023133\relax
\mciteBstWouldAddEndPuncttrue
\mciteSetBstMidEndSepPunct{\mcitedefaultmidpunct}
{\mcitedefaultendpunct}{\mcitedefaultseppunct}\relax
\EndOfBibitem
\bibitem[Garcia-Perez \latin{et~al.}(2020)Garcia-Perez, Rossi, and
  Maniscalco]{Garcia-Perez2020}
Garcia-Perez,~G.; Rossi,~M. A.~C.; Maniscalco,~S. {IBM Q E}xperience as a
  versatile experimental testbed for simulating open quantum systems. \emph{NPJ
  Quantum Inf.} \textbf{2020}, \emph{6}, 1\relax
\mciteBstWouldAddEndPuncttrue
\mciteSetBstMidEndSepPunct{\mcitedefaultmidpunct}
{\mcitedefaultendpunct}{\mcitedefaultseppunct}\relax
\EndOfBibitem
\bibitem[Endo \latin{et~al.}(2020)Endo, Sun, Li, Benjamin, and Yuan]{Endo2020}
Endo,~S.; Sun,~J.; Li,~Y.; Benjamin,~S.~C.; Yuan,~X. Variational Quantum
  Simulation of General Processes. \emph{Phys. Rev. Lett.} \textbf{2020},
  \emph{125}, 010501\relax
\mciteBstWouldAddEndPuncttrue
\mciteSetBstMidEndSepPunct{\mcitedefaultmidpunct}
{\mcitedefaultendpunct}{\mcitedefaultseppunct}\relax
\EndOfBibitem
\bibitem[Schlimgen \latin{et~al.}(2021)Schlimgen, Head-Marsden, Sager, Narang,
  and Mazziotti]{Schlimgen2021}
Schlimgen,~A.~W.; Head-Marsden,~K.; Sager,~L.~M.; Narang,~P.; Mazziotti,~D.~A.
  Quantum Simulation of Open Quantum Systems Using a Unitary Decomposition of
  Operators. \emph{Phys. Rev. Lett.} \textbf{2021}, \emph{127}, 270503\relax
\mciteBstWouldAddEndPuncttrue
\mciteSetBstMidEndSepPunct{\mcitedefaultmidpunct}
{\mcitedefaultendpunct}{\mcitedefaultseppunct}\relax
\EndOfBibitem
\bibitem[Kamakari \latin{et~al.}(2022)Kamakari, Sun, Motta, and
  Minnich]{Kamakari2022}
Kamakari,~H.; Sun,~S.-N.; Motta,~M.; Minnich,~A.~J. Digital Quantum Simulation
  of Open Quantum Systems Using Quantum Imaginary--Time Evolution. \emph{PRX
  Quantum} \textbf{2022}, \emph{3}, 010320\relax
\mciteBstWouldAddEndPuncttrue
\mciteSetBstMidEndSepPunct{\mcitedefaultmidpunct}
{\mcitedefaultendpunct}{\mcitedefaultseppunct}\relax
\EndOfBibitem
\bibitem[Suri \latin{et~al.}(2023)Suri, Barreto, Hadfield, Wiebe, Wudarski, and
  Marshall]{Suri2023}
Suri,~N.; Barreto,~J.; Hadfield,~S.; Wiebe,~N.; Wudarski,~F.; Marshall,~J.
  Two-Unitary Decomposition Algorithm and Open Quantum System Simulation.
  \emph{Quantum} \textbf{2023}, \emph{7}, 1002\relax
\mciteBstWouldAddEndPuncttrue
\mciteSetBstMidEndSepPunct{\mcitedefaultmidpunct}
{\mcitedefaultendpunct}{\mcitedefaultseppunct}\relax
\EndOfBibitem
\bibitem[Schlimgen \latin{et~al.}(2022)Schlimgen, Head-Marsden, Sager-Smith,
  Narang, and Mazziotti]{Schlimgen2022a}
Schlimgen,~A.~W.; Head-Marsden,~K.; Sager-Smith,~L.~M.; Narang,~P.;
  Mazziotti,~D.~A. Quantum state preparation and nonunitary evolution with
  diagonal operators. \emph{Physical Review A} \textbf{2022}, \emph{106}\relax
\mciteBstWouldAddEndPuncttrue
\mciteSetBstMidEndSepPunct{\mcitedefaultmidpunct}
{\mcitedefaultendpunct}{\mcitedefaultseppunct}\relax
\EndOfBibitem
\bibitem[Gaikwad \latin{et~al.}(2022)Gaikwad, Arvind, and Dorai]{Gaikwad2022}
Gaikwad,~A.; Arvind,; Dorai,~K. Simulating open quantum dynamics on an NMR
  quantum processor using the Sz.-Nagy dilation algorithm. \emph{Phys. Rev. A}
  \textbf{2022}, \emph{106}, 022424\relax
\mciteBstWouldAddEndPuncttrue
\mciteSetBstMidEndSepPunct{\mcitedefaultmidpunct}
{\mcitedefaultendpunct}{\mcitedefaultseppunct}\relax
\EndOfBibitem
\bibitem[Ding \latin{et~al.}(2024)Ding, Li, and Lin]{Ding2024}
Ding,~Z.; Li,~X.; Lin,~L. {Simulating Open Quantum Systems Using Hamiltonian
  Simulations}. \emph{PRX Quantum} \textbf{2024}, \emph{5}, 020332\relax
\mciteBstWouldAddEndPuncttrue
\mciteSetBstMidEndSepPunct{\mcitedefaultmidpunct}
{\mcitedefaultendpunct}{\mcitedefaultseppunct}\relax
\EndOfBibitem
\bibitem[Basile and Pineda(2024)Basile, and Pineda]{Basile2024}
Basile,~T.; Pineda,~C. {Quantum simulation of Pauli channels and dynamical
  maps: Algorithm and implementation}. \emph{PLOS One} \textbf{2024},
  \emph{19}, e0297210\relax
\mciteBstWouldAddEndPuncttrue
\mciteSetBstMidEndSepPunct{\mcitedefaultmidpunct}
{\mcitedefaultendpunct}{\mcitedefaultseppunct}\relax
\EndOfBibitem
\bibitem[Xuereb \latin{et~al.}(2023)Xuereb, Campbell, Goold, and
  Xuereb]{Xuereb2023}
Xuereb,~J.; Campbell,~S.; Goold,~J.; Xuereb,~A. Deterministic quantum
  computation with one-clean-qubit model as an open quantum system.
  \emph{Physical Review A} \textbf{2023}, \emph{107}\relax
\mciteBstWouldAddEndPuncttrue
\mciteSetBstMidEndSepPunct{\mcitedefaultmidpunct}
{\mcitedefaultendpunct}{\mcitedefaultseppunct}\relax
\EndOfBibitem
\bibitem[Shivpuje \latin{et~al.}(2024)Shivpuje, Sajjan, Wang, Hu, and
  Kais]{Shivpuje2024}
Shivpuje,~S.; Sajjan,~M.; Wang,~Y.; Hu,~Z.; Kais,~S. Designing Variational
  Ansatz for Quantum-Enabled Simulation of Non-Unitary Dynamical Evolution - An
  Excursion into Dicke Supperradiance. \emph{Advanced Quantum Technologies}
  \textbf{2024}, \relax
\mciteBstWouldAddEndPunctfalse
\mciteSetBstMidEndSepPunct{\mcitedefaultmidpunct}
{}{\mcitedefaultseppunct}\relax
\EndOfBibitem
\bibitem[Watad and Lindner(2024)Watad, and Lindner]{Watad2024}
Watad,~T.~M.; Lindner,~N.~H. Variational quantum algorithms for simulation of
  Lindblad dynamics. \emph{Quantum Science and Technology} \textbf{2024},
  \emph{9}\relax
\mciteBstWouldAddEndPuncttrue
\mciteSetBstMidEndSepPunct{\mcitedefaultmidpunct}
{\mcitedefaultendpunct}{\mcitedefaultseppunct}\relax
\EndOfBibitem
\bibitem[Luo \latin{et~al.}(2024)Luo, Lin, and Gao]{Luo2024}
Luo,~J.; Lin,~K.; Gao,~X. Variational Quantum Simulation of Lindblad Dynamics
  via Quantum State Diffusion. \emph{Journal OF Physical Chemistry Letters}
  \textbf{2024}, \emph{15}, 3516--3522\relax
\mciteBstWouldAddEndPuncttrue
\mciteSetBstMidEndSepPunct{\mcitedefaultmidpunct}
{\mcitedefaultendpunct}{\mcitedefaultseppunct}\relax
\EndOfBibitem
\bibitem[Mahdian and Yeganeh(2020)Mahdian, and Yeganeh]{Mahdian2020a}
Mahdian,~M.; Yeganeh,~H.~D. Hybrid quantum variational algorithm for simulating
  open quantum systems with near-term devices. \emph{Journal of Physics A -
  Mathematical and Theoretical} \textbf{2020}, \emph{53}\relax
\mciteBstWouldAddEndPuncttrue
\mciteSetBstMidEndSepPunct{\mcitedefaultmidpunct}
{\mcitedefaultendpunct}{\mcitedefaultseppunct}\relax
\EndOfBibitem
\bibitem[Liu \latin{et~al.}(2021)Liu, Sun, Wu, and Guo]{Liu2021}
Liu,~H.-Y.; Sun,~T.-P.; Wu,~Y.-C.; Guo,~G.-P. Variational Quantum Algorithms
  for the Steady States of Open Quantum Systems. \emph{Chinese Physics Letters}
  \textbf{2021}, \emph{38}\relax
\mciteBstWouldAddEndPuncttrue
\mciteSetBstMidEndSepPunct{\mcitedefaultmidpunct}
{\mcitedefaultendpunct}{\mcitedefaultseppunct}\relax
\EndOfBibitem
\bibitem[Lau \latin{et~al.}(2023)Lau, Lim, Bharti, Kwek, and
  Vinjanampathy]{Lau2023}
Lau,~J. W.~Z.; Lim,~K.~H.; Bharti,~K.; Kwek,~L.-C.; Vinjanampathy,~S. {Convex
  Optimization for Nonequilibrium Steady States on a Hybrid Quantum Processor}.
  \emph{Physical Review Letters} \textbf{2023}, \emph{130}, 240601\relax
\mciteBstWouldAddEndPuncttrue
\mciteSetBstMidEndSepPunct{\mcitedefaultmidpunct}
{\mcitedefaultendpunct}{\mcitedefaultseppunct}\relax
\EndOfBibitem
\bibitem[Joo and Spiller(2023)Joo, and Spiller]{Joo2023}
Joo,~J.; Spiller,~T.~P. Commutation simulator for open quantum dynamics.
  \emph{New Journal of Physics} \textbf{2023}, \emph{25}\relax
\mciteBstWouldAddEndPuncttrue
\mciteSetBstMidEndSepPunct{\mcitedefaultmidpunct}
{\mcitedefaultendpunct}{\mcitedefaultseppunct}\relax
\EndOfBibitem
\bibitem[Suri \latin{et~al.}(2018)Suri, Binder, Muralidharan, and
  Vinjanampathy]{Suri2018}
Suri,~N.; Binder,~F.~C.; Muralidharan,~B.; Vinjanampathy,~S. {Speeding up
  thermalisation via open quantum system variational optimisation}. \emph{The
  European Physical Journal Special Topics} \textbf{2018}, \emph{227},
  203--216\relax
\mciteBstWouldAddEndPuncttrue
\mciteSetBstMidEndSepPunct{\mcitedefaultmidpunct}
{\mcitedefaultendpunct}{\mcitedefaultseppunct}\relax
\EndOfBibitem
\bibitem[Santos \latin{et~al.}(2024)Santos, Song, and Savona]{Santos2024}
Santos,~S.; Song,~X.; Savona,~V. {Low-Rank Variational Quantum Algorithm for
  the Dynamics of Open Quantum Systems}. \emph{arXiv} \textbf{2024}, \relax
\mciteBstWouldAddEndPunctfalse
\mciteSetBstMidEndSepPunct{\mcitedefaultmidpunct}
{}{\mcitedefaultseppunct}\relax
\EndOfBibitem
\bibitem[Lee \latin{et~al.}(2021)Lee, Patil, Zhang, and Hsieh]{Lee2021}
Lee,~C.~K.; Patil,~P.; Zhang,~S.; Hsieh,~C.~Y. {Neural-network variational
  quantum algorithm for simulating many-body dynamics}. \emph{Physical Review
  Research} \textbf{2021}, \emph{3}, 023095\relax
\mciteBstWouldAddEndPuncttrue
\mciteSetBstMidEndSepPunct{\mcitedefaultmidpunct}
{\mcitedefaultendpunct}{\mcitedefaultseppunct}\relax
\EndOfBibitem
\bibitem[Ollitrault \latin{et~al.}(2023)Ollitrault, Jandura, Miessen,
  Burghardt, Martinazzo, Tacchino, and Tavernelli]{Ollitrault2023}
Ollitrault,~P.~J.; Jandura,~S.; Miessen,~A.; Burghardt,~I.; Martinazzo,~R.;
  Tacchino,~F.; Tavernelli,~I. Quantum algorithms for grid-based variational
  time evolution. \emph{Quantum} \textbf{2023}, \emph{7}\relax
\mciteBstWouldAddEndPuncttrue
\mciteSetBstMidEndSepPunct{\mcitedefaultmidpunct}
{\mcitedefaultendpunct}{\mcitedefaultseppunct}\relax
\EndOfBibitem
\bibitem[Gravina and Savona(2024)Gravina, and Savona]{Gravina2024}
Gravina,~L.; Savona,~V. Adaptive variational low-rank dynamics for open quantum
  systems. \emph{Phys. Rev. Res.} \textbf{2024}, \emph{6}, 023072\relax
\mciteBstWouldAddEndPuncttrue
\mciteSetBstMidEndSepPunct{\mcitedefaultmidpunct}
{\mcitedefaultendpunct}{\mcitedefaultseppunct}\relax
\EndOfBibitem
\bibitem[Schlegel \latin{et~al.}(2023)Schlegel, Minganti, and
  Savona]{Schlegel2023}
Schlegel,~D.~S.; Minganti,~F.; Savona,~V. {Coherent-State Ladder Time-Dependent
  Variational Principle for Open Quantum Systems}. \emph{arXiv} \textbf{2023},
  \relax
\mciteBstWouldAddEndPunctfalse
\mciteSetBstMidEndSepPunct{\mcitedefaultmidpunct}
{}{\mcitedefaultseppunct}\relax
\EndOfBibitem
\bibitem[Zhou \latin{et~al.}(2023)Zhou, Mao, and Sun]{Zhou2023}
Zhou,~H.; Mao,~R.; Sun,~X. {Hybrid algorithm simulating non-equilibrium steady
  states of an open quantum system}. \emph{arXiv} \textbf{2023}, \relax
\mciteBstWouldAddEndPunctfalse
\mciteSetBstMidEndSepPunct{\mcitedefaultmidpunct}
{}{\mcitedefaultseppunct}\relax
\EndOfBibitem
\bibitem[Lyu \latin{et~al.}(2023)Lyu, Mulvihill, Soley, Geva, and
  Batista]{Lyu2023}
Lyu,~N.; Mulvihill,~E.; Soley,~M.~B.; Geva,~E.; Batista,~V.~S. {Tensor-Train
  Thermo-Field Memory Kernels for Generalized Quantum Master Equations}.
  \emph{Journal of Chemical Theory and Computation} \textbf{2023}, \emph{19},
  1111--1129\relax
\mciteBstWouldAddEndPuncttrue
\mciteSetBstMidEndSepPunct{\mcitedefaultmidpunct}
{\mcitedefaultendpunct}{\mcitedefaultseppunct}\relax
\EndOfBibitem
\bibitem[Gupta and Chandrashekar(2020)Gupta, and Chandrashekar]{Gupta2020b}
Gupta,~P.; Chandrashekar,~C.~M. {Optimal quantum simulation of open quantum
  systems}. \emph{arXiv} \textbf{2020}, \relax
\mciteBstWouldAddEndPunctfalse
\mciteSetBstMidEndSepPunct{\mcitedefaultmidpunct}
{}{\mcitedefaultseppunct}\relax
\EndOfBibitem
\bibitem[Barreiro \latin{et~al.}(2011)Barreiro, Mueller, Schindler, Nigg, Monz,
  Chwalla, Hennrich, Roos, Zoller, and Blatt]{Berreiro2011}
Barreiro,~J.~T.; Mueller,~M.; Schindler,~P.; Nigg,~D.; Monz,~T.; Chwalla,~M.;
  Hennrich,~M.; Roos,~C.~F.; Zoller,~P.; Blatt,~R. An open-system quantum
  simulator with trapped ions. \emph{Nature} \textbf{2011}, \emph{470},
  486--491\relax
\mciteBstWouldAddEndPuncttrue
\mciteSetBstMidEndSepPunct{\mcitedefaultmidpunct}
{\mcitedefaultendpunct}{\mcitedefaultseppunct}\relax
\EndOfBibitem
\bibitem[Mostame \latin{et~al.}(2017)Mostame, Huh, Kreisbeck, Kerman, Fujita,
  Eisfeld, and Aspuru-Guzik]{Mostame2017}
Mostame,~S.; Huh,~J.; Kreisbeck,~C.; Kerman,~A.~J.; Fujita,~T.; Eisfeld,~A.;
  Aspuru-Guzik,~A. Emulation of complex open quantum systems using
  superconducting qubits. \emph{Quantum Information Processing} \textbf{2017},
  \emph{16}\relax
\mciteBstWouldAddEndPuncttrue
\mciteSetBstMidEndSepPunct{\mcitedefaultmidpunct}
{\mcitedefaultendpunct}{\mcitedefaultseppunct}\relax
\EndOfBibitem
\bibitem[Su and Li(2020)Su, and Li]{Su2020}
Su,~H.-Y.; Li,~Y. Quantum algorithm for the simulation of open-system dynamics
  and thermalization. \emph{Phys. Rev. A} \textbf{2020}, \emph{101},
  012328\relax
\mciteBstWouldAddEndPuncttrue
\mciteSetBstMidEndSepPunct{\mcitedefaultmidpunct}
{\mcitedefaultendpunct}{\mcitedefaultseppunct}\relax
\EndOfBibitem
\bibitem[Sun \latin{et~al.}(2021)Sun, Shih, and Cheng]{Sun2021}
Sun,~S.; Shih,~L.-C.; Cheng,~Y.-C. Efficient Quantum Simulation of Open Quantum
  System Dynamics on Noisy Quantum Computers. 2021;
  \url{https://arxiv.org/abs/2106.12882}\relax
\mciteBstWouldAddEndPuncttrue
\mciteSetBstMidEndSepPunct{\mcitedefaultmidpunct}
{\mcitedefaultendpunct}{\mcitedefaultseppunct}\relax
\EndOfBibitem
\bibitem[Burger \latin{et~al.}(2022)Burger, Kwek, and Poletti]{Burger2022}
Burger,~A.; Kwek,~L.~C.; Poletti,~D. Digital Quantum Simulation of the
  Spin-Boson Model under Markovian Open-System Dynamics. \emph{Entropy}
  \textbf{2022}, \emph{24}\relax
\mciteBstWouldAddEndPuncttrue
\mciteSetBstMidEndSepPunct{\mcitedefaultmidpunct}
{\mcitedefaultendpunct}{\mcitedefaultseppunct}\relax
\EndOfBibitem
\bibitem[Wang \latin{et~al.}(2022)Wang, Zhang, and Li]{Wang2022}
Wang,~A.; Zhang,~J.; Li,~Y. Error-mitigated deep-circuit quantum simulation of
  open systems: Steady state and relaxation rate problems. \emph{Physical
  Review Research} \textbf{2022}, \emph{4}\relax
\mciteBstWouldAddEndPuncttrue
\mciteSetBstMidEndSepPunct{\mcitedefaultmidpunct}
{\mcitedefaultendpunct}{\mcitedefaultseppunct}\relax
\EndOfBibitem
\bibitem[Cattaneo \latin{et~al.}(2023)Cattaneo, Rossi, Garcia-Perez, Zambrini,
  and Maniscalco]{Cattaneo2023}
Cattaneo,~M.; Rossi,~M. A.~C.; Garcia-Perez,~G.; Zambrini,~R.; Maniscalco,~S.
  Quantum Simulation of Dissipative Collective Effects on Noisy Quantum
  Computers. \emph{PRX Quantum} \textbf{2023}, \emph{4}\relax
\mciteBstWouldAddEndPuncttrue
\mciteSetBstMidEndSepPunct{\mcitedefaultmidpunct}
{\mcitedefaultendpunct}{\mcitedefaultseppunct}\relax
\EndOfBibitem
\bibitem[Zanetti \latin{et~al.}(2023)Zanetti, Pinto, Basso, and
  Maziero]{Zanetti2023}
Zanetti,~M.~S.; Pinto,~D.~F.; Basso,~M. L.~W.; Maziero,~J. Simulating noisy
  quantum channels via quantum state preparation algorithms. \emph{Journal of
  Physics B-Atomic Molecular and Optical Physics} \textbf{2023},
  \emph{56}\relax
\mciteBstWouldAddEndPuncttrue
\mciteSetBstMidEndSepPunct{\mcitedefaultmidpunct}
{\mcitedefaultendpunct}{\mcitedefaultseppunct}\relax
\EndOfBibitem
\bibitem[Leppaekangas \latin{et~al.}(2023)Leppaekangas, Vogt, Fratus, Bark,
  Vaitkus, Stadler, Reiner, Zanker, and Marthaler]{Leppaekangas2023}
Leppaekangas,~J.; Vogt,~N.; Fratus,~K.~R.; Bark,~K.; Vaitkus,~J.~A.;
  Stadler,~P.; Reiner,~J.-M.; Zanker,~S.; Marthaler,~M. Quantum algorithm for
  solving open-system dynamics on quantum computers using noise. \emph{Physical
  Review A} \textbf{2023}, \emph{108}\relax
\mciteBstWouldAddEndPuncttrue
\mciteSetBstMidEndSepPunct{\mcitedefaultmidpunct}
{\mcitedefaultendpunct}{\mcitedefaultseppunct}\relax
\EndOfBibitem
\bibitem[Guimaraes \latin{et~al.}(2023)Guimaraes, Lim, Vasilevskiy, Huelga, and
  Plenio]{Guimaraes2023}
Guimaraes,~J.~D.; Lim,~J.; Vasilevskiy,~M.~I.; Huelga,~S.~F.; Plenio,~M.~B.
  Noise-Assisted Digital Quantum Simulation of Open Systems Using Partial
  Probabilistic Error Cancellation. \emph{PRX Quantum} \textbf{2023},
  \emph{4}\relax
\mciteBstWouldAddEndPuncttrue
\mciteSetBstMidEndSepPunct{\mcitedefaultmidpunct}
{\mcitedefaultendpunct}{\mcitedefaultseppunct}\relax
\EndOfBibitem
\bibitem[Guimaraes \latin{et~al.}(2024)Guimaraes, Vasilevskiy, and
  Barbosa]{Guimaraes2024}
Guimaraes,~J.~D.; Vasilevskiy,~M.~I.; Barbosa,~L.~S. Digital quantum simulation
  of non-perturbative dynamics of open systems with orthogonal polynomials.
  \emph{Quantum} \textbf{2024}, \emph{8}\relax
\mciteBstWouldAddEndPuncttrue
\mciteSetBstMidEndSepPunct{\mcitedefaultmidpunct}
{\mcitedefaultendpunct}{\mcitedefaultseppunct}\relax
\EndOfBibitem
\bibitem[Del~Re \latin{et~al.}(2024)Del~Re, Rost, Foss-Feig, Kemper, and
  Freericks]{DelRe2024}
Del~Re,~L.; Rost,~B.; Foss-Feig,~M.; Kemper,~A.~F.; Freericks,~J.~K. Robust
  Measurements of n-Point Correlation Functions of Driven-Dissipative Quantum
  Systems on a Digital Quantum Computer. \emph{Physical Review Letters}
  \textbf{2024}, \emph{132}\relax
\mciteBstWouldAddEndPuncttrue
\mciteSetBstMidEndSepPunct{\mcitedefaultmidpunct}
{\mcitedefaultendpunct}{\mcitedefaultseppunct}\relax
\EndOfBibitem
\bibitem[Childs and Li(2017)Childs, and Li]{Childs2017}
Childs,~A.~M.; Li,~T. Efficient Simulation of Sparse Markovian Quantum
  Dynamics. \emph{Quantum Info. Comput.} \textbf{2017}, \emph{17},
  901–947\relax
\mciteBstWouldAddEndPuncttrue
\mciteSetBstMidEndSepPunct{\mcitedefaultmidpunct}
{\mcitedefaultendpunct}{\mcitedefaultseppunct}\relax
\EndOfBibitem
\bibitem[Li and Wang(2023)Li, and Wang]{Li2023}
Li,~X.; Wang,~C. {Succinct Description and Efficient Simulation of
  Non-Markovian Open Quantum Systems}. \emph{Communications in Mathematical
  Physics} \textbf{2023}, \emph{401}, 147--183\relax
\mciteBstWouldAddEndPuncttrue
\mciteSetBstMidEndSepPunct{\mcitedefaultmidpunct}
{\mcitedefaultendpunct}{\mcitedefaultseppunct}\relax
\EndOfBibitem
\bibitem[Li and Wang(2023)Li, and Wang]{Li2023a}
Li,~X.; Wang,~C. Simulating Markovian open quantum systems using higher-order
  series expansion. \textbf{2023}, \relax
\mciteBstWouldAddEndPunctfalse
\mciteSetBstMidEndSepPunct{\mcitedefaultmidpunct}
{}{\mcitedefaultseppunct}\relax
\EndOfBibitem
\bibitem[Cosacchi \latin{et~al.}(2018)Cosacchi, Cygorek, Ungar, Barth, Vagov,
  and Axt]{Cosacchi2018}
Cosacchi,~M.; Cygorek,~M.; Ungar,~F.; Barth,~A.~M.; Vagov,~A.; Axt,~V.~M.
  Path-integral approach for nonequilibrium multitime correlation functions of
  open quantum systems coupled to Markovian and non-Markovian environments.
  \emph{Physical Review B} \textbf{2018}, \emph{98}\relax
\mciteBstWouldAddEndPuncttrue
\mciteSetBstMidEndSepPunct{\mcitedefaultmidpunct}
{\mcitedefaultendpunct}{\mcitedefaultseppunct}\relax
\EndOfBibitem
\bibitem[Liu \latin{et~al.}(2019)Liu, Segal, and Hanna]{Liu2019}
Liu,~J.; Segal,~D.; Hanna,~G. Hybrid quantum-classical simulation of quantum
  speed limits in open quantum systems. \emph{Journal of Physics A-Mathematical
  and Theoretical} \textbf{2019}, \emph{52}\relax
\mciteBstWouldAddEndPuncttrue
\mciteSetBstMidEndSepPunct{\mcitedefaultmidpunct}
{\mcitedefaultendpunct}{\mcitedefaultseppunct}\relax
\EndOfBibitem
\bibitem[Cygorek \latin{et~al.}(2022)Cygorek, Cosacchi, Vagov, Axt, Lovett,
  Keeling, and Gauger]{Cygorek2022}
Cygorek,~M.; Cosacchi,~M.; Vagov,~A.; Axt,~V.~M.; Lovett,~B.~W.; Keeling,~J.;
  Gauger,~E.~M. Simulation of open quantum systems by automated compression of
  arbitrary environments. \emph{Nature Physics} \textbf{2022}, \emph{18},
  662\relax
\mciteBstWouldAddEndPuncttrue
\mciteSetBstMidEndSepPunct{\mcitedefaultmidpunct}
{\mcitedefaultendpunct}{\mcitedefaultseppunct}\relax
\EndOfBibitem
\bibitem[Andreadakis \latin{et~al.}(2023)Andreadakis, Anand, and
  Zanardi]{Andreadakis2023}
Andreadakis,~F.; Anand,~N.; Zanardi,~P. Scrambling of algebras in open quantum
  systems. \emph{Physical Review A} \textbf{2023}, \emph{107}\relax
\mciteBstWouldAddEndPuncttrue
\mciteSetBstMidEndSepPunct{\mcitedefaultmidpunct}
{\mcitedefaultendpunct}{\mcitedefaultseppunct}\relax
\EndOfBibitem
\bibitem[R\'{e}gent and Rouchon(2023)R\'{e}gent, and Rouchon]{Regent2023}
R\'{e}gent,~F.-M.~L.; Rouchon,~P. {Adiabatic elimination for composite open
  quantum systems: reduced model formulation and numerical simulations}.
  \emph{arXiv} \textbf{2023}, \relax
\mciteBstWouldAddEndPunctfalse
\mciteSetBstMidEndSepPunct{\mcitedefaultmidpunct}
{}{\mcitedefaultseppunct}\relax
\EndOfBibitem
\bibitem[M\"{u}ller \latin{et~al.}(2011)M\"{u}ller, Hammerer, Zhou, Roos, and
  Zoller]{Muller2011}
M\"{u}ller,~M.; Hammerer,~K.; Zhou,~Y.~L.; Roos,~C.~F.; Zoller,~P. {Simulating
  open quantum systems: from many-body interactions to stabilizer pumping}.
  \emph{New Journal of Physics} \textbf{2011}, \emph{13}, 085007\relax
\mciteBstWouldAddEndPuncttrue
\mciteSetBstMidEndSepPunct{\mcitedefaultmidpunct}
{\mcitedefaultendpunct}{\mcitedefaultseppunct}\relax
\EndOfBibitem
\bibitem[Ramusat and Savona(2021)Ramusat, and Savona]{Ramusat2021}
Ramusat,~N.; Savona,~V. A quantum algorithm for the direct estimation of the
  steady state of open quantum systems. \emph{Quantum} \textbf{2021},
  \emph{5}\relax
\mciteBstWouldAddEndPuncttrue
\mciteSetBstMidEndSepPunct{\mcitedefaultmidpunct}
{\mcitedefaultendpunct}{\mcitedefaultseppunct}\relax
\EndOfBibitem
\bibitem[Mahdian and Yeganeh(2020)Mahdian, and Yeganeh]{Mahdian2020b}
Mahdian,~M.; Yeganeh,~H.~D. Incoherent quantum algorithm dynamics of an open
  system with near-term devices. \emph{Quantum Information Processing}
  \textbf{2020}, \emph{19}\relax
\mciteBstWouldAddEndPuncttrue
\mciteSetBstMidEndSepPunct{\mcitedefaultmidpunct}
{\mcitedefaultendpunct}{\mcitedefaultseppunct}\relax
\EndOfBibitem
\bibitem[Rost \latin{et~al.}(2020)Rost, Jones, Vyushkova, Ali, Cullip,
  Vyushkov, and Nabrzyski]{Rost2020}
Rost,~B.; Jones,~B.; Vyushkova,~M.; Ali,~A.; Cullip,~C.; Vyushkov,~A.;
  Nabrzyski,~J. Simulation of Thermal Relaxation in Spin Chemistry Systems on a
  Quantum Computer Using Inherent Qubit Decoherence. 2020;
  \url{https://arxiv.org/abs/2001.00794}\relax
\mciteBstWouldAddEndPuncttrue
\mciteSetBstMidEndSepPunct{\mcitedefaultmidpunct}
{\mcitedefaultendpunct}{\mcitedefaultseppunct}\relax
\EndOfBibitem
\bibitem[Tolunay \latin{et~al.}(2023)Tolunay, Liepuoniute, Vyushkova, and
  Jones]{Tolunay2023}
Tolunay,~M.; Liepuoniute,~I.; Vyushkova,~M.; Jones,~B.~A. Hamiltonian
  simulation of quantum beats in radical pairs undergoing thermal relaxation on
  near-term quantum computers. \emph{Phys. Chem. Chem. Phys.} \textbf{2023},
  \emph{25}, 15115--15134\relax
\mciteBstWouldAddEndPuncttrue
\mciteSetBstMidEndSepPunct{\mcitedefaultmidpunct}
{\mcitedefaultendpunct}{\mcitedefaultseppunct}\relax
\EndOfBibitem
\bibitem[Mostame \latin{et~al.}(2012)Mostame, Rebentrost, Eisfeld, Kerman,
  Tsomokos, and Aspuru-Guzik]{Mostame2012}
Mostame,~S.; Rebentrost,~P.; Eisfeld,~A.; Kerman,~A.~J.; Tsomokos,~D.~I.;
  Aspuru-Guzik,~A. Quantum simulator of an open quantum system using
  superconducting qubits: exciton transport in photosynthetic complexes.
  \emph{New Journal of Physics} \textbf{2012}, \emph{14}\relax
\mciteBstWouldAddEndPuncttrue
\mciteSetBstMidEndSepPunct{\mcitedefaultmidpunct}
{\mcitedefaultendpunct}{\mcitedefaultseppunct}\relax
\EndOfBibitem
\bibitem[Gupta and Chandrashekar(2020)Gupta, and Chandrashekar]{Gupta2020}
Gupta,~P.; Chandrashekar,~C.~M. Digital quantum simulation framework for energy
  transport in an open quantum system. \emph{New Journal of Physics}
  \textbf{2020}, \emph{22}\relax
\mciteBstWouldAddEndPuncttrue
\mciteSetBstMidEndSepPunct{\mcitedefaultmidpunct}
{\mcitedefaultendpunct}{\mcitedefaultseppunct}\relax
\EndOfBibitem
\bibitem[Hu \latin{et~al.}(2022)Hu, Head-Marsden, Mazziotti, Narang, and
  Kais]{Hu2022}
Hu,~Z.; Head-Marsden,~K.; Mazziotti,~D.~A.; Narang,~P.; Kais,~S. A general
  quantum algorithm for open quantum dynamics demonstrated with the
  {F}enna-{M}atthews-{O}lson complex. \emph{{Quantum}} \textbf{2022}, \emph{6},
  726\relax
\mciteBstWouldAddEndPuncttrue
\mciteSetBstMidEndSepPunct{\mcitedefaultmidpunct}
{\mcitedefaultendpunct}{\mcitedefaultseppunct}\relax
\EndOfBibitem
\bibitem[Zhang \latin{et~al.}(2023)Zhang, Hu, Wang, and Kais]{Zhang2023}
Zhang,~Y.; Hu,~Z.; Wang,~Y.; Kais,~S. Quantum Simulation of the Radical Pair
  Dynamics of the Avian Compass. \emph{The Journal of Physical Chemistry
  Letters} \textbf{2023}, \emph{14}, 832--837\relax
\mciteBstWouldAddEndPuncttrue
\mciteSetBstMidEndSepPunct{\mcitedefaultmidpunct}
{\mcitedefaultendpunct}{\mcitedefaultseppunct}\relax
\EndOfBibitem
\bibitem[Oh \latin{et~al.}(0)Oh, Krogmeier, Schlimgen, and
  Head-Marsden]{Oh2024}
Oh,~E.~K.; Krogmeier,~T.~J.; Schlimgen,~A.~W.; Head-Marsden,~K. Singular Value
  Decomposition Quantum Algorithm for Quantum Biology. \emph{ACS Physical
  Chemistry Au} \textbf{0}, \emph{0}, null\relax
\mciteBstWouldAddEndPuncttrue
\mciteSetBstMidEndSepPunct{\mcitedefaultmidpunct}
{\mcitedefaultendpunct}{\mcitedefaultseppunct}\relax
\EndOfBibitem
\bibitem[Sun \latin{et~al.}(2024)Sun, Shih, and Cheng]{Sun2024}
Sun,~S.; Shih,~L.-C.; Cheng,~Y.-C. Efficient quantum simulation of open quantum
  system dynamics on noisy quantum computers. \emph{Physica Scripta}
  \textbf{2024}, \emph{99}\relax
\mciteBstWouldAddEndPuncttrue
\mciteSetBstMidEndSepPunct{\mcitedefaultmidpunct}
{\mcitedefaultendpunct}{\mcitedefaultseppunct}\relax
\EndOfBibitem
\bibitem[Rost \latin{et~al.}(2021)Rost, Del~Re, Earnest, Kemper, Jones, and
  Freericks]{Rost2021}
Rost,~B.; Del~Re,~L.; Earnest,~N.; Kemper,~A.~F.; Jones,~B.; Freericks,~J.~K.
  Demonstrating robust simulation of driven-dissipative problems on near-term
  quantum computers. 2021; \url{https://arxiv.org/abs/2108.01183}\relax
\mciteBstWouldAddEndPuncttrue
\mciteSetBstMidEndSepPunct{\mcitedefaultmidpunct}
{\mcitedefaultendpunct}{\mcitedefaultseppunct}\relax
\EndOfBibitem
\bibitem[Schlimgen \latin{et~al.}(2022)Schlimgen, Head-Marsden, Sager, Narang,
  and Mazziotti]{Schlimgen2022b}
Schlimgen,~A.~W.; Head-Marsden,~K.; Sager,~L.~M.; Narang,~P.; Mazziotti,~D.~A.
  Quantum simulation of the Lindblad equation using a unitary decomposition of
  operators. \emph{Phys. Rev. Res.} \textbf{2022}, \emph{4}, 023216\relax
\mciteBstWouldAddEndPuncttrue
\mciteSetBstMidEndSepPunct{\mcitedefaultmidpunct}
{\mcitedefaultendpunct}{\mcitedefaultseppunct}\relax
\EndOfBibitem
\bibitem[Tornow \latin{et~al.}(2022)Tornow, Gehrke, and Helmbrecht]{Tornow2022}
Tornow,~S.; Gehrke,~W.; Helmbrecht,~U. {Non-equilibrium dynamics of a
  dissipative two-site Hubbard model simulated on IBM quantum computers}.
  \emph{Journal of Physics A: Mathematical and Theoretical} \textbf{2022},
  \emph{55}, 245302\relax
\mciteBstWouldAddEndPuncttrue
\mciteSetBstMidEndSepPunct{\mcitedefaultmidpunct}
{\mcitedefaultendpunct}{\mcitedefaultseppunct}\relax
\EndOfBibitem
\bibitem[Tan \latin{et~al.}(2023)Tan, Sun, Tazhigulov, Chan, and
  Minnich]{Tan2023}
Tan,~A. T.~K.; Sun,~S.-N.; Tazhigulov,~R.~N.; Chan,~G. K.-L.; Minnich,~A.~J.
  Realizing symmetry-protected topological phases in a spin-1/2 chain with
  next-nearest-neighbor hopping on superconducting qubits. \emph{Phycial Review
  A} \textbf{2023}, \emph{107}\relax
\mciteBstWouldAddEndPuncttrue
\mciteSetBstMidEndSepPunct{\mcitedefaultmidpunct}
{\mcitedefaultendpunct}{\mcitedefaultseppunct}\relax
\EndOfBibitem
\bibitem[Liang \latin{et~al.}(2023)Liang, Lv, Wang, and Fei]{Liang2023}
Liang,~J.-M.; Lv,~Q.-Q.; Wang,~Z.-X.; Fei,~S.-M. Assisted quantum simulation of
  open quantum systems. \emph{IScience} \textbf{2023}, \emph{26}\relax
\mciteBstWouldAddEndPuncttrue
\mciteSetBstMidEndSepPunct{\mcitedefaultmidpunct}
{\mcitedefaultendpunct}{\mcitedefaultseppunct}\relax
\EndOfBibitem
\bibitem[Jong \latin{et~al.}(2022)Jong, Lee, Mulligan, Płoskoń, Ringer, and
  Yao]{Jong2022}
Jong,~W. A.~d.; Lee,~K.; Mulligan,~J.; Płoskoń,~M.; Ringer,~F.; Yao,~X.
  {Quantum simulation of nonequilibrium dynamics and thermalization in the
  Schwinger model}. \emph{Physical Review D} \textbf{2022}, \emph{106},
  054508\relax
\mciteBstWouldAddEndPuncttrue
\mciteSetBstMidEndSepPunct{\mcitedefaultmidpunct}
{\mcitedefaultendpunct}{\mcitedefaultseppunct}\relax
\EndOfBibitem
\bibitem[Del~Re \latin{et~al.}(2020)Del~Re, Rost, Kemper, and
  Freericks]{DelRe2020}
Del~Re,~L.; Rost,~B.; Kemper,~A.~F.; Freericks,~J.~K. Driven-dissipative
  quantum mechanics on a lattice: Simulating a fermionic reservoir on a quantum
  computer. \emph{Phys. Rev. B} \textbf{2020}, \emph{102}, 125112\relax
\mciteBstWouldAddEndPuncttrue
\mciteSetBstMidEndSepPunct{\mcitedefaultmidpunct}
{\mcitedefaultendpunct}{\mcitedefaultseppunct}\relax
\EndOfBibitem
\bibitem[Head-Marsden \latin{et~al.}(2021)Head-Marsden, Krastanov, Mazziotti,
  and Narang]{Head-Marsden2021}
Head-Marsden,~K.; Krastanov,~S.; Mazziotti,~D.~A.; Narang,~P. Capturing
  non-Markovian dynamics on near-term quantum computers. \emph{Phys. Rev. Res.}
  \textbf{2021}, \emph{3}, 013182\relax
\mciteBstWouldAddEndPuncttrue
\mciteSetBstMidEndSepPunct{\mcitedefaultmidpunct}
{\mcitedefaultendpunct}{\mcitedefaultseppunct}\relax
\EndOfBibitem
\bibitem[Warren \latin{et~al.}(2024)Warren, Wang, Benavides-Riveros, and
  Mazziotti]{Warren2024}
Warren,~S.; Wang,~Y.; Benavides-Riveros,~C.~L.; Mazziotti,~D.~A. {Exact Ansatz
  of Fermion-Boson Systems for a Quantum Device}. \emph{arXiv} \textbf{2024},
  \relax
\mciteBstWouldAddEndPunctfalse
\mciteSetBstMidEndSepPunct{\mcitedefaultmidpunct}
{}{\mcitedefaultseppunct}\relax
\EndOfBibitem
\bibitem[Wang \latin{et~al.}(2023)Wang, Mulvihill, Hu, Lyu, Shivpuje, Liu,
  Soley, Geva, Batista, and Kais]{Wang2023}
Wang,~Y.; Mulvihill,~E.; Hu,~Z.; Lyu,~N.; Shivpuje,~S.; Liu,~Y.; Soley,~M.~B.;
  Geva,~E.; Batista,~V.~S.; Kais,~S. Simulating Open Quantum System Dynamics on
  NISQ Computers with Generalized Quantum Master Equations. \emph{Journal of
  Chemical Theory and Computation} \textbf{2023}, \emph{19}, 4851--4862\relax
\mciteBstWouldAddEndPuncttrue
\mciteSetBstMidEndSepPunct{\mcitedefaultmidpunct}
{\mcitedefaultendpunct}{\mcitedefaultseppunct}\relax
\EndOfBibitem
\bibitem[Jagadish and Petruccione(2018)Jagadish, and Petruccione]{Jagadish2019}
Jagadish,~V.; Petruccione,~F. An Invitation to Quantum Channels. \emph{Quanta}
  \textbf{2018}, \emph{7}, 54--67\relax
\mciteBstWouldAddEndPuncttrue
\mciteSetBstMidEndSepPunct{\mcitedefaultmidpunct}
{\mcitedefaultendpunct}{\mcitedefaultseppunct}\relax
\EndOfBibitem
\bibitem[Pechukas(1994)]{Pechukas1994}
Pechukas,~P. Reduced Dynamics Need Not Be Completely Positive. \emph{Phys. Rev.
  Lett.} \textbf{1994}, \emph{73}, 1060--1062\relax
\mciteBstWouldAddEndPuncttrue
\mciteSetBstMidEndSepPunct{\mcitedefaultmidpunct}
{\mcitedefaultendpunct}{\mcitedefaultseppunct}\relax
\EndOfBibitem
\bibitem[Carteret \latin{et~al.}(2008)Carteret, Terno, and
  \ifmmode~\dot{Z}\else \.{Z}\fi{}yczkowski]{Carteret2008}
Carteret,~H.~A.; Terno,~D.~R.; \ifmmode~\dot{Z}\else \.{Z}\fi{}yczkowski,~K.
  Dynamics beyond completely positive maps: Some properties and applications.
  \emph{Phys. Rev. A} \textbf{2008}, \emph{77}, 042113\relax
\mciteBstWouldAddEndPuncttrue
\mciteSetBstMidEndSepPunct{\mcitedefaultmidpunct}
{\mcitedefaultendpunct}{\mcitedefaultseppunct}\relax
\EndOfBibitem
\bibitem[Sargolzahi(2020)]{Sargolzahi2020}
Sargolzahi,~I. Positivity of the assignment map implies complete positivity of
  the reduced dynamics. \emph{Quantum Information Processing} \textbf{2020},
  \emph{19}, 310\relax
\mciteBstWouldAddEndPuncttrue
\mciteSetBstMidEndSepPunct{\mcitedefaultmidpunct}
{\mcitedefaultendpunct}{\mcitedefaultseppunct}\relax
\EndOfBibitem
\bibitem[Manzano(2020)]{Manzano2020}
Manzano,~D. {A short introduction to the Lindblad master equation}. \emph{AIP
  Advances} \textbf{2020}, \emph{10}, 025106\relax
\mciteBstWouldAddEndPuncttrue
\mciteSetBstMidEndSepPunct{\mcitedefaultmidpunct}
{\mcitedefaultendpunct}{\mcitedefaultseppunct}\relax
\EndOfBibitem
\bibitem[Alicki(1995)]{Alicki1995}
Alicki,~R. Comment on ``Reduced Dynamics Need Not Be Completely Positive''.
  \emph{Phys. Rev. Lett.} \textbf{1995}, \emph{75}, 3020--3020\relax
\mciteBstWouldAddEndPuncttrue
\mciteSetBstMidEndSepPunct{\mcitedefaultmidpunct}
{\mcitedefaultendpunct}{\mcitedefaultseppunct}\relax
\EndOfBibitem
\bibitem[Pechukas(1995)]{Pechukas1995}
Pechukas,~P. Pechukas Replies. \emph{Phys. Rev. Lett.} \textbf{1995},
  \emph{75}, 3021--3021\relax
\mciteBstWouldAddEndPuncttrue
\mciteSetBstMidEndSepPunct{\mcitedefaultmidpunct}
{\mcitedefaultendpunct}{\mcitedefaultseppunct}\relax
\EndOfBibitem
\bibitem[Levy and Shalit(2014)Levy, and Shalit]{Levy2014}
Levy,~E.; Shalit,~O.~M. {Dilation theory in finite dimensions: The possible,
  the impossible and the unknown}. \emph{Rocky Mountain Journal of Mathematics}
  \textbf{2014}, \emph{44}, 203 -- 221\relax
\mciteBstWouldAddEndPuncttrue
\mciteSetBstMidEndSepPunct{\mcitedefaultmidpunct}
{\mcitedefaultendpunct}{\mcitedefaultseppunct}\relax
\EndOfBibitem
\bibitem[Ticozzi and Viola(2017)Ticozzi, and Viola]{Ticozzi2017}
Ticozzi,~F.; Viola,~L. Quantum and classical resources for unitary design of
  open-system evolutions. \emph{Quantum Science and Technology} \textbf{2017},
  \emph{2}, 034001\relax
\mciteBstWouldAddEndPuncttrue
\mciteSetBstMidEndSepPunct{\mcitedefaultmidpunct}
{\mcitedefaultendpunct}{\mcitedefaultseppunct}\relax
\EndOfBibitem
\bibitem[Havel(2003)]{Havel2003}
Havel,~T.~F. Robust procedures for converting among Lindblad, Kraus and matrix
  representations of quantum dynamical semigroups. \emph{J. Math. Phys.}
  \textbf{2003}, \emph{44}, 534--557\relax
\mciteBstWouldAddEndPuncttrue
\mciteSetBstMidEndSepPunct{\mcitedefaultmidpunct}
{\mcitedefaultendpunct}{\mcitedefaultseppunct}\relax
\EndOfBibitem
\bibitem[Kunold(2024)]{Kunold2024}
Kunold,~A. Vectorization of the density matrix and quantum simulation of the
  von Neumann equation of time-dependent Hamiltonians. \emph{Physica Scripta}
  \textbf{2024}, \emph{99}\relax
\mciteBstWouldAddEndPuncttrue
\mciteSetBstMidEndSepPunct{\mcitedefaultmidpunct}
{\mcitedefaultendpunct}{\mcitedefaultseppunct}\relax
\EndOfBibitem
\bibitem[Byrd \latin{et~al.}(2016)Byrd, Ceballos, and Chitambar]{Byrd2016}
Byrd,~M.; Ceballos,~R.; Chitambar,~E. Open system quantum evolution and the
  assumption of complete positivity (A Tutorial). \emph{International Journal
  of Quantum Information} \textbf{2016}, \emph{14}\relax
\mciteBstWouldAddEndPuncttrue
\mciteSetBstMidEndSepPunct{\mcitedefaultmidpunct}
{\mcitedefaultendpunct}{\mcitedefaultseppunct}\relax
\EndOfBibitem
\bibitem[Alicki and Lendi(2007)Alicki, and Lendi]{Alicki2007}
Alicki,~R.; Lendi,~K. \emph{Quantum dynamical semigroups and applications};
  Springer, 2007; Vol. 717\relax
\mciteBstWouldAddEndPuncttrue
\mciteSetBstMidEndSepPunct{\mcitedefaultmidpunct}
{\mcitedefaultendpunct}{\mcitedefaultseppunct}\relax
\EndOfBibitem
\bibitem[Cubitt \latin{et~al.}(2012)Cubitt, Eisert, and Wolf]{Cubitt2012}
Cubitt,~T.~S.; Eisert,~J.; Wolf,~M.~M. The Complexity of Relating Quantum
  Channels to Master Equations. \emph{Communcations in Mathematical Physics}
  \textbf{2012}, \emph{310}, 383--418\relax
\mciteBstWouldAddEndPuncttrue
\mciteSetBstMidEndSepPunct{\mcitedefaultmidpunct}
{\mcitedefaultendpunct}{\mcitedefaultseppunct}\relax
\EndOfBibitem
\bibitem[Gorini \latin{et~al.}(1976)Gorini, Kossakowski, and
  Sudarshan]{Gorini1976}
Gorini,~V.; Kossakowski,~A.; Sudarshan,~E. C.~G. {Completely positive dynamical
  semigroups of N‐level systems}. \emph{Journal of Mathematical Physics}
  \textbf{1976}, \emph{17}, 821--825\relax
\mciteBstWouldAddEndPuncttrue
\mciteSetBstMidEndSepPunct{\mcitedefaultmidpunct}
{\mcitedefaultendpunct}{\mcitedefaultseppunct}\relax
\EndOfBibitem
\bibitem[Lindblad(1976)]{Lindblad1976}
Lindblad,~G. On the generators of quantum dynamical semigroups.
  \emph{Communications in Mathematical Physics} \textbf{1976}, \emph{48},
  119--130\relax
\mciteBstWouldAddEndPuncttrue
\mciteSetBstMidEndSepPunct{\mcitedefaultmidpunct}
{\mcitedefaultendpunct}{\mcitedefaultseppunct}\relax
\EndOfBibitem
\bibitem[Gorini \latin{et~al.}(2008)Gorini, Kossakowski, and
  Sudarshan]{Gorini2008}
Gorini,~V.; Kossakowski,~A.; Sudarshan,~E. C.~G. {Completely positive dynamical
  semigroups of N‐level systems}. \emph{Journal of Mathematical Physics}
  \textbf{2008}, \emph{17}, 821--825\relax
\mciteBstWouldAddEndPuncttrue
\mciteSetBstMidEndSepPunct{\mcitedefaultmidpunct}
{\mcitedefaultendpunct}{\mcitedefaultseppunct}\relax
\EndOfBibitem
\bibitem[Redfield(1965)]{Redfield1965}
Redfield,~A.~G. In \emph{Advances in Magnetic Resonance}; Waugh,~J.~S., Ed.;
  Advances in Magnetic and Optical Resonance; Academic Press, 1965; Vol.~1; pp
  1--32\relax
\mciteBstWouldAddEndPuncttrue
\mciteSetBstMidEndSepPunct{\mcitedefaultmidpunct}
{\mcitedefaultendpunct}{\mcitedefaultseppunct}\relax
\EndOfBibitem
\bibitem[Nakajima(1958)]{Nakajima1958}
Nakajima,~S. On quantum theory of transport phenomena: steady diffusion.
  \emph{Progress of Theoretical Physics} \textbf{1958}, \emph{20},
  948--959\relax
\mciteBstWouldAddEndPuncttrue
\mciteSetBstMidEndSepPunct{\mcitedefaultmidpunct}
{\mcitedefaultendpunct}{\mcitedefaultseppunct}\relax
\EndOfBibitem
\bibitem[Zwanzig(1960)]{Zwanzig1960}
Zwanzig,~R. Ensemble method in the theory of irreversibility. \emph{The Journal
  of Chemical Physics} \textbf{1960}, \emph{33}, 1338--1341\relax
\mciteBstWouldAddEndPuncttrue
\mciteSetBstMidEndSepPunct{\mcitedefaultmidpunct}
{\mcitedefaultendpunct}{\mcitedefaultseppunct}\relax
\EndOfBibitem
\bibitem[Haake(1973)]{Haake1973}
Haake,~F. \emph{Springer Tracts in Modern Physics: Ergebnisse der exakten
  Naturwissenschaftenc; Volume 66}; Springer, 1973; pp 98--168\relax
\mciteBstWouldAddEndPuncttrue
\mciteSetBstMidEndSepPunct{\mcitedefaultmidpunct}
{\mcitedefaultendpunct}{\mcitedefaultseppunct}\relax
\EndOfBibitem
\bibitem[Tokuyama and Mori(1976)Tokuyama, and Mori]{Tokuyama1976}
Tokuyama,~M.; Mori,~H. {Statistical-Mechanical Theory of Random Frequency
  Modulations and Generalized Brownian Motions*)}. \emph{Progress of
  Theoretical Physics} \textbf{1976}, \emph{55}, 411--429\relax
\mciteBstWouldAddEndPuncttrue
\mciteSetBstMidEndSepPunct{\mcitedefaultmidpunct}
{\mcitedefaultendpunct}{\mcitedefaultseppunct}\relax
\EndOfBibitem
\bibitem[Shibata \latin{et~al.}(1977)Shibata, Takahashi, and
  Hashitsume]{Shibata1977}
Shibata,~F.; Takahashi,~Y.; Hashitsume,~N. A generalized stochastic liouville
  equation. Non-Markovian versus memoryless master equations. \emph{Journal of
  Statistical Physics} \textbf{1977}, \emph{17}, 171--187\relax
\mciteBstWouldAddEndPuncttrue
\mciteSetBstMidEndSepPunct{\mcitedefaultmidpunct}
{\mcitedefaultendpunct}{\mcitedefaultseppunct}\relax
\EndOfBibitem
\bibitem[de~Vega and Alonso(2017)de~Vega, and Alonso]{deVega2017}
de~Vega,~I.; Alonso,~D. Dynamics of non-Markovian open quantum systems.
  \emph{Reviews of Modern Physics} \textbf{2017}, \emph{89}\relax
\mciteBstWouldAddEndPuncttrue
\mciteSetBstMidEndSepPunct{\mcitedefaultmidpunct}
{\mcitedefaultendpunct}{\mcitedefaultseppunct}\relax
\EndOfBibitem
\bibitem[Sweke \latin{et~al.}(2014)Sweke, Sinayskiy, and
  Petruccione]{Sweke2014}
Sweke,~R.; Sinayskiy,~I.; Petruccione,~F. Simulation of single-qubit open
  quantum systems. \emph{Phys. Rev. A} \textbf{2014}, \emph{90}, 022331\relax
\mciteBstWouldAddEndPuncttrue
\mciteSetBstMidEndSepPunct{\mcitedefaultmidpunct}
{\mcitedefaultendpunct}{\mcitedefaultseppunct}\relax
\EndOfBibitem
\bibitem[David \latin{et~al.}(2023)David, Sinayskiy, and
  Petruccione]{David2023}
David,~I.; Sinayskiy,~I.; Petruccione,~F. Digital Simulation of Single Qubit
  Markovian Open Quantum Systems: A Tutorial. \emph{Quanta} \textbf{2023},
  \emph{12}, 131--163\relax
\mciteBstWouldAddEndPuncttrue
\mciteSetBstMidEndSepPunct{\mcitedefaultmidpunct}
{\mcitedefaultendpunct}{\mcitedefaultseppunct}\relax
\EndOfBibitem
\bibitem[Buscemi \latin{et~al.}(2003)Buscemi, D'Ariano, and
  Sacchi]{Buscemi2003}
Buscemi,~F.; D'Ariano,~G.~M.; Sacchi,~M.~F. Physical realizations of quantum
  operations. \emph{Phys. Rev. A} \textbf{2003}, \emph{68}, 042113\relax
\mciteBstWouldAddEndPuncttrue
\mciteSetBstMidEndSepPunct{\mcitedefaultmidpunct}
{\mcitedefaultendpunct}{\mcitedefaultseppunct}\relax
\EndOfBibitem
\bibitem[Childs and Wiebe(2012)Childs, and Wiebe]{Childs2012}
Childs,~A.~M.; Wiebe,~N. Hamiltonian simulation using linear combinations of
  unitary operations. \emph{Quantum Info. Comput.} \textbf{2012}, \emph{12},
  901–924\relax
\mciteBstWouldAddEndPuncttrue
\mciteSetBstMidEndSepPunct{\mcitedefaultmidpunct}
{\mcitedefaultendpunct}{\mcitedefaultseppunct}\relax
\EndOfBibitem
\bibitem[Smart and Mazziotti(2024)Smart, and Mazziotti]{Smart2024}
Smart,~S.~E.; Mazziotti,~D.~A. {Verifiably exact solution of the electronic
  Schrödinger equation on quantum devices}. \emph{Physical Review A}
  \textbf{2024}, \emph{109}, 022802\relax
\mciteBstWouldAddEndPuncttrue
\mciteSetBstMidEndSepPunct{\mcitedefaultmidpunct}
{\mcitedefaultendpunct}{\mcitedefaultseppunct}\relax
\EndOfBibitem
\bibitem[Wang and Mazziotti(2023)Wang, and Mazziotti]{Wang2023b}
Wang,~Y.; Mazziotti,~D.~A. {Electronic excited states from a variance-based
  contracted quantum eigensolver}. \emph{Physical Review A} \textbf{2023},
  \emph{108}, 022814\relax
\mciteBstWouldAddEndPuncttrue
\mciteSetBstMidEndSepPunct{\mcitedefaultmidpunct}
{\mcitedefaultendpunct}{\mcitedefaultseppunct}\relax
\EndOfBibitem
\bibitem[Smart \latin{et~al.}(2022)Smart, Boyn, and Mazziotti]{Smart2022}
Smart,~S.~E.; Boyn,~J.-N.; Mazziotti,~D.~A. Resolving correlated states of
  benzyne with an error-mitigated contracted quantum eigensolver.
  \emph{Physical Review A} \textbf{2022}, \emph{105}\relax
\mciteBstWouldAddEndPuncttrue
\mciteSetBstMidEndSepPunct{\mcitedefaultmidpunct}
{\mcitedefaultendpunct}{\mcitedefaultseppunct}\relax
\EndOfBibitem
\bibitem[Smart and Mazziotti(2021)Smart, and Mazziotti]{Smart2021}
Smart,~S.~E.; Mazziotti,~D.~A. Quantum solver of contracted eigenvalue
  equations for scalable molecular simulations on quantum computing devices.
  \emph{Physical Review Letters} \textbf{2021}, \emph{126}, 070504\relax
\mciteBstWouldAddEndPuncttrue
\mciteSetBstMidEndSepPunct{\mcitedefaultmidpunct}
{\mcitedefaultendpunct}{\mcitedefaultseppunct}\relax
\EndOfBibitem
\bibitem[Mazzola(2024)]{Mazzola2024}
Mazzola,~G. Quantum computing for chemistry and physics applications from a
  Monte Carlo perspective. \emph{Journal of Chemical Physics} \textbf{2024},
  \emph{160}\relax
\mciteBstWouldAddEndPuncttrue
\mciteSetBstMidEndSepPunct{\mcitedefaultmidpunct}
{\mcitedefaultendpunct}{\mcitedefaultseppunct}\relax
\EndOfBibitem
\bibitem[Nagy and Savona(2018)Nagy, and Savona]{Nagy2018}
Nagy,~A.; Savona,~V. Driven-dissipative quantum Monte Carlo method for open
  quantum systems. \emph{Physical Review A} \textbf{2018}, \emph{97}\relax
\mciteBstWouldAddEndPuncttrue
\mciteSetBstMidEndSepPunct{\mcitedefaultmidpunct}
{\mcitedefaultendpunct}{\mcitedefaultseppunct}\relax
\EndOfBibitem
\bibitem[Kadowaki(2018)]{Kadowaki2018}
Kadowaki,~T. Dynamics of open quantum systems by interpolation of von Neumann
  and classical master equations, and its application to quantum annealing.
  \emph{Physical Review A} \textbf{2018}, \emph{97}\relax
\mciteBstWouldAddEndPuncttrue
\mciteSetBstMidEndSepPunct{\mcitedefaultmidpunct}
{\mcitedefaultendpunct}{\mcitedefaultseppunct}\relax
\EndOfBibitem
\bibitem[Peetz \latin{et~al.}(2023)Peetz, Smart, Tserkis, and
  Narang]{Peetz2023}
Peetz,~J.; Smart,~S.~E.; Tserkis,~S.; Narang,~P. Simulation of {{Open Quantum
  Systems}} via {{Low-Depth Convex Unitary Evolutions}}.
  https://arxiv.org/abs/2307.14325v2, 2023\relax
\mciteBstWouldAddEndPuncttrue
\mciteSetBstMidEndSepPunct{\mcitedefaultmidpunct}
{\mcitedefaultendpunct}{\mcitedefaultseppunct}\relax
\EndOfBibitem
\bibitem[Rosgen(2008)]{Rosgen2008}
Rosgen,~B. Additivity and {{Distinguishability}} of {{Random Unitary
  Channels}}. \emph{Journal of Mathematical Physics} \textbf{2008}, \emph{49},
  102107\relax
\mciteBstWouldAddEndPuncttrue
\mciteSetBstMidEndSepPunct{\mcitedefaultmidpunct}
{\mcitedefaultendpunct}{\mcitedefaultseppunct}\relax
\EndOfBibitem
\bibitem[Schlimgen \latin{et~al.}(2022)Schlimgen, Head-Marsden, Sager-Smith,
  Narang, and Mazziotti]{Schlimgen2022}
Schlimgen,~A.~W.; Head-Marsden,~K.; Sager-Smith,~L.~M.; Narang,~P.;
  Mazziotti,~D.~A. Quantum Simulation of Open Quantum Systems Using
  Density-Matrix Purification. 2022\relax
\mciteBstWouldAddEndPuncttrue
\mciteSetBstMidEndSepPunct{\mcitedefaultmidpunct}
{\mcitedefaultendpunct}{\mcitedefaultseppunct}\relax
\EndOfBibitem
\bibitem[Delgado-Granados \latin{et~al.}(2024)Delgado-Granados, Warren, and
  Mazziotti]{Delgado-Granados2024}
Delgado-Granados,~L.~H.; Warren,~S.; Mazziotti,~D.~A. Unitary Dynamics for Open
  Quantum Systems with Density-Matrix Purification. 2024\relax
\mciteBstWouldAddEndPuncttrue
\mciteSetBstMidEndSepPunct{\mcitedefaultmidpunct}
{\mcitedefaultendpunct}{\mcitedefaultseppunct}\relax
\EndOfBibitem
\bibitem[Zong \latin{et~al.}(2024)Zong, Huai, Cai, Jin, Zhan, Zhang, Bu, Sui,
  Fei, Zheng, Zhang, Wu, and Yin]{Zong2024}
Zong,~Z.; Huai,~S.; Cai,~T.; Jin,~W.; Zhan,~Z.; Zhang,~Z.; Bu,~K.; Sui,~L.;
  Fei,~Y.; Zheng,~Y.; Zhang,~S.; Wu,~J.; Yin,~Y. Determination of molecular
  energies via variational-based quantum imaginary time evolution in a
  superconducting qubit system. \emph{Science China-Physics Mechanics \&
  Astronomy} \textbf{2024}, \emph{67}\relax
\mciteBstWouldAddEndPuncttrue
\mciteSetBstMidEndSepPunct{\mcitedefaultmidpunct}
{\mcitedefaultendpunct}{\mcitedefaultseppunct}\relax
\EndOfBibitem
\bibitem[Kleinmann \latin{et~al.}(2006)Kleinmann, Kampermann, Meyer, and
  Bru\ss{}]{Kleinmann2006}
Kleinmann,~M.; Kampermann,~H.; Meyer,~T.; Bru\ss{},~D. Physical purification of
  quantum states. \emph{Phys. Rev. A} \textbf{2006}, \emph{73}, 062309\relax
\mciteBstWouldAddEndPuncttrue
\mciteSetBstMidEndSepPunct{\mcitedefaultmidpunct}
{\mcitedefaultendpunct}{\mcitedefaultseppunct}\relax
\EndOfBibitem
\bibitem[Bassi and Ghirardi(2003)Bassi, and Ghirardi]{Bassi2003}
Bassi,~A.; Ghirardi,~G. A general scheme for ensemble purification.
  \emph{Physics Letters A} \textbf{2003}, \emph{309}, 24--28\relax
\mciteBstWouldAddEndPuncttrue
\mciteSetBstMidEndSepPunct{\mcitedefaultmidpunct}
{\mcitedefaultendpunct}{\mcitedefaultseppunct}\relax
\EndOfBibitem
\bibitem[Hughston \latin{et~al.}(1993)Hughston, Jozsa, and
  Wootters]{Hughston1993}
Hughston,~L.~P.; Jozsa,~R.; Wootters,~W.~K. A complete classification of
  quantum ensembles having a given density matrix. \emph{Physics Letters A}
  \textbf{1993}, \emph{183}, 14--18\relax
\mciteBstWouldAddEndPuncttrue
\mciteSetBstMidEndSepPunct{\mcitedefaultmidpunct}
{\mcitedefaultendpunct}{\mcitedefaultseppunct}\relax
\EndOfBibitem
\bibitem[Motta \latin{et~al.}(2020)Motta, Sun, Tan, O'Rourke, Ye, Minnich,
  Brandão, and Chan]{Motta2020}
Motta,~M.; Sun,~C.; Tan,~A. T.~K.; O'Rourke,~M.~J.; Ye,~E.; Minnich,~A.~J.;
  Brandão,~F. G. S.~L.; Chan,~G. K.-L. Determining eigenstates and thermal
  states on a quantum computer using quantum imaginary time evolution.
  \emph{Nature Physics} \textbf{2020}, \emph{16}, 205--210\relax
\mciteBstWouldAddEndPuncttrue
\mciteSetBstMidEndSepPunct{\mcitedefaultmidpunct}
{\mcitedefaultendpunct}{\mcitedefaultseppunct}\relax
\EndOfBibitem
\bibitem[Cao \latin{et~al.}(2022)Cao, An, Hou, Zhou, and Zeng]{Cao2022}
Cao,~C.; An,~Z.; Hou,~S.-Y.; Zhou,~D.~L.; Zeng,~B. Quantum imaginary time
  evolution steered by reinforcement learning. \emph{Communications Physics}
  \textbf{2022}, \emph{5}\relax
\mciteBstWouldAddEndPuncttrue
\mciteSetBstMidEndSepPunct{\mcitedefaultmidpunct}
{\mcitedefaultendpunct}{\mcitedefaultseppunct}\relax
\EndOfBibitem
\bibitem[Mao \latin{et~al.}(2023)Mao, Chaudhary, Kondappan, Shi, Ilo-Okeke,
  Ivannikov, and Byrnes]{Mao2023}
Mao,~Y.; Chaudhary,~M.; Kondappan,~M.; Shi,~J.; Ilo-Okeke,~E.~O.;
  Ivannikov,~V.; Byrnes,~T. Measurement-Based Deterministic Imaginary Time
  Evolution. \emph{Physical Review Letters} \textbf{2023}, \emph{131}\relax
\mciteBstWouldAddEndPuncttrue
\mciteSetBstMidEndSepPunct{\mcitedefaultmidpunct}
{\mcitedefaultendpunct}{\mcitedefaultseppunct}\relax
\EndOfBibitem
\bibitem[Lin \latin{et~al.}(2021)Lin, Dilip, Green, Smith, and
  Pollmann]{Lin2021}
Lin,~S.-H.; Dilip,~R.; Green,~A.~G.; Smith,~A.; Pollmann,~F. {Real- and
  Imaginary-Time Evolution with Compressed Quantum Circuits}. \emph{PRX
  Quantum} \textbf{2021}, \emph{2}\relax
\mciteBstWouldAddEndPuncttrue
\mciteSetBstMidEndSepPunct{\mcitedefaultmidpunct}
{\mcitedefaultendpunct}{\mcitedefaultseppunct}\relax
\EndOfBibitem
\bibitem[Inoue and Fukumoto(2018)Inoue, and Fukumoto]{Inoue2018}
Inoue,~K.; Fukumoto,~Y. {Typical Purification Reproducing the Time Evolution of
  an Open Quantum System}. \emph{arXiv} \textbf{2018}, \relax
\mciteBstWouldAddEndPunctfalse
\mciteSetBstMidEndSepPunct{\mcitedefaultmidpunct}
{}{\mcitedefaultseppunct}\relax
\EndOfBibitem
\bibitem[McArdle \latin{et~al.}(2019)McArdle, Jones, Endo, Li, Benjamin, and
  Yuan]{McArdle2019}
McArdle,~S.; Jones,~T.; Endo,~S.; Li,~Y.; Benjamin,~S.~C.; Yuan,~X. Variational
  ansatz-based quantum simulation of imaginary time evolution. \emph{npj
  Quantum Information} \textbf{2019}, \emph{5}, 75\relax
\mciteBstWouldAddEndPuncttrue
\mciteSetBstMidEndSepPunct{\mcitedefaultmidpunct}
{\mcitedefaultendpunct}{\mcitedefaultseppunct}\relax
\EndOfBibitem
\bibitem[Shirakawa \latin{et~al.}(2021)Shirakawa, Seki, and
  Yunoki]{Shirakawa2021}
Shirakawa,~T.; Seki,~K.; Yunoki,~S. Discretized quantum adiabatic process for
  free fermions and comparison with the imaginary-time evolution. \emph{Phys.
  Rev. Res.} \textbf{2021}, \emph{3}, 013004\relax
\mciteBstWouldAddEndPuncttrue
\mciteSetBstMidEndSepPunct{\mcitedefaultmidpunct}
{\mcitedefaultendpunct}{\mcitedefaultseppunct}\relax
\EndOfBibitem
\bibitem[Nishi \latin{et~al.}(2021)Nishi, Kosugi, and
  Matsushita]{Matsushita2021}
Nishi,~H.; Kosugi,~T.; Matsushita,~Y.-i. Implementation of quantum
  imaginary-time evolution method on NISQ devices by introducing nonlocal
  approximation. \emph{npj Quantum Information} \textbf{2021}, \emph{7},
  85\relax
\mciteBstWouldAddEndPuncttrue
\mciteSetBstMidEndSepPunct{\mcitedefaultmidpunct}
{\mcitedefaultendpunct}{\mcitedefaultseppunct}\relax
\EndOfBibitem
\bibitem[Lin \latin{et~al.}(2021)Lin, Dilip, Green, Smith, and
  Pollmann]{Pollmann2021}
Lin,~S.-H.; Dilip,~R.; Green,~A.~G.; Smith,~A.; Pollmann,~F. Real- and
  Imaginary-Time Evolution with Compressed Quantum Circuits. \emph{PRX Quantum}
  \textbf{2021}, \emph{2}, 010342\relax
\mciteBstWouldAddEndPuncttrue
\mciteSetBstMidEndSepPunct{\mcitedefaultmidpunct}
{\mcitedefaultendpunct}{\mcitedefaultseppunct}\relax
\EndOfBibitem
\bibitem[Kramer and Saraceno(1981)Kramer, and Saraceno]{Kramer1981}
Kramer,~P.; Saraceno,~M. \emph{Geometry of the time-dependent variational
  principle in quantum mechanics}; Springer, 1981\relax
\mciteBstWouldAddEndPuncttrue
\mciteSetBstMidEndSepPunct{\mcitedefaultmidpunct}
{\mcitedefaultendpunct}{\mcitedefaultseppunct}\relax
\EndOfBibitem
\bibitem[Broeckhove \latin{et~al.}(1988)Broeckhove, Lathouwers, Kesteloot, and
  Van~Leuven]{Broeckhove1988}
Broeckhove,~J.; Lathouwers,~L.; Kesteloot,~E.; Van~Leuven,~P. On the
  equivalence of time-dependent variational principles. \emph{Chemical physics
  letters} \textbf{1988}, \emph{149}, 547--550\relax
\mciteBstWouldAddEndPuncttrue
\mciteSetBstMidEndSepPunct{\mcitedefaultmidpunct}
{\mcitedefaultendpunct}{\mcitedefaultseppunct}\relax
\EndOfBibitem
\bibitem[Pedersen and Koch(1998)Pedersen, and Koch]{Pedersen1998}
Pedersen,~T.~B.; Koch,~H. On the time-dependent Lagrangian approach in quantum
  chemistry. \emph{The Journal of chemical physics} \textbf{1998}, \emph{108},
  5194--5204\relax
\mciteBstWouldAddEndPuncttrue
\mciteSetBstMidEndSepPunct{\mcitedefaultmidpunct}
{\mcitedefaultendpunct}{\mcitedefaultseppunct}\relax
\EndOfBibitem
\bibitem[Kerman and Koonin(1976)Kerman, and Koonin]{Kerman1976}
Kerman,~A.~K.; Koonin,~S.~E. Hamiltonian formulation of time-dependent
  variational principles for the many-body system. \emph{Annals of Physics}
  \textbf{1976}, \emph{100}, 332--358\relax
\mciteBstWouldAddEndPuncttrue
\mciteSetBstMidEndSepPunct{\mcitedefaultmidpunct}
{\mcitedefaultendpunct}{\mcitedefaultseppunct}\relax
\EndOfBibitem
\bibitem[Yuan \latin{et~al.}(2019)Yuan, Endo, Zhao, Li, and Benjamin]{Yuan2019}
Yuan,~X.; Endo,~S.; Zhao,~Q.; Li,~Y.; Benjamin,~S.~C. Theory of variational
  quantum simulation. \emph{Quantum} \textbf{2019}, \emph{3}, 191\relax
\mciteBstWouldAddEndPuncttrue
\mciteSetBstMidEndSepPunct{\mcitedefaultmidpunct}
{\mcitedefaultendpunct}{\mcitedefaultseppunct}\relax
\EndOfBibitem
\bibitem[Hackl \latin{et~al.}(2020)Hackl, Guaita, Shi, Haegeman, Demler, and
  Cirac]{Hackl2020}
Hackl,~L.; Guaita,~T.; Shi,~T.; Haegeman,~J.; Demler,~E.; Cirac,~J.~I. Geometry
  of variational methods: dynamics of closed quantum systems. \emph{SciPost
  Physics} \textbf{2020}, \emph{9}, 048\relax
\mciteBstWouldAddEndPuncttrue
\mciteSetBstMidEndSepPunct{\mcitedefaultmidpunct}
{\mcitedefaultendpunct}{\mcitedefaultseppunct}\relax
\EndOfBibitem
\bibitem[Haegeman \latin{et~al.}(2013)Haegeman, Osborne, and
  Verstraete]{Haegeman2013}
Haegeman,~J.; Osborne,~T.~J.; Verstraete,~F. Post-matrix product state methods:
  To tangent space and beyond. \emph{Phys. Rev. B} \textbf{2013}, \emph{88},
  075133\relax
\mciteBstWouldAddEndPuncttrue
\mciteSetBstMidEndSepPunct{\mcitedefaultmidpunct}
{\mcitedefaultendpunct}{\mcitedefaultseppunct}\relax
\EndOfBibitem
\bibitem[Dirac(1930)]{Dirac1930}
Dirac,~P. A.~M. Note on exchange phenomena in the Thomas atom.
  \emph{Mathematical Proceedings of the Cambridge Philosophical Society}
  \textbf{1930}, \emph{26}, 376–385\relax
\mciteBstWouldAddEndPuncttrue
\mciteSetBstMidEndSepPunct{\mcitedefaultmidpunct}
{\mcitedefaultendpunct}{\mcitedefaultseppunct}\relax
\EndOfBibitem
\bibitem[Frenkel(1934)]{Frenkel1934}
Frenkel,~J. \emph{Wave Mechanics: Advanced General Theory}; 1934\relax
\mciteBstWouldAddEndPuncttrue
\mciteSetBstMidEndSepPunct{\mcitedefaultmidpunct}
{\mcitedefaultendpunct}{\mcitedefaultseppunct}\relax
\EndOfBibitem
\bibitem[Langhoff \latin{et~al.}(1972)Langhoff, Epstein, and
  Karplus]{Langhoff1972}
Langhoff,~P.; Epstein,~S.; Karplus,~M. Aspects of time-dependent perturbation
  theory. \emph{Reviews of Modern Physics} \textbf{1972}, \emph{44}, 602\relax
\mciteBstWouldAddEndPuncttrue
\mciteSetBstMidEndSepPunct{\mcitedefaultmidpunct}
{\mcitedefaultendpunct}{\mcitedefaultseppunct}\relax
\EndOfBibitem
\bibitem[McLachlan(1964)]{McLachlan1964}
McLachlan,~A. A variational solution of the time-dependent Schrodinger
  equation. \emph{Molecular Physics} \textbf{1964}, \emph{8}, 39--44\relax
\mciteBstWouldAddEndPuncttrue
\mciteSetBstMidEndSepPunct{\mcitedefaultmidpunct}
{\mcitedefaultendpunct}{\mcitedefaultseppunct}\relax
\EndOfBibitem
\bibitem[Messina \latin{et~al.}(1994)Messina, Garrett, and
  Schenter]{Messina1994}
Messina,~M.; Garrett,~B.~C.; Schenter,~G.~K. Variational solutions for the
  thermal and real time propagator using the McLachlan variational principle.
  \emph{The Journal of chemical physics} \textbf{1994}, \emph{100},
  6570--6577\relax
\mciteBstWouldAddEndPuncttrue
\mciteSetBstMidEndSepPunct{\mcitedefaultmidpunct}
{\mcitedefaultendpunct}{\mcitedefaultseppunct}\relax
\EndOfBibitem
\bibitem[Yao \latin{et~al.}(2021)Yao, Gomes, Zhang, Wang, Ho, Iadecola, and
  Orth]{Yao2021}
Yao,~Y.-X.; Gomes,~N.; Zhang,~F.; Wang,~C.-Z.; Ho,~K.-M.; Iadecola,~T.;
  Orth,~P.~P. Adaptive variational quantum dynamics simulations. \emph{PRX
  Quantum} \textbf{2021}, \emph{2}, 030307\relax
\mciteBstWouldAddEndPuncttrue
\mciteSetBstMidEndSepPunct{\mcitedefaultmidpunct}
{\mcitedefaultendpunct}{\mcitedefaultseppunct}\relax
\EndOfBibitem
\bibitem[Lee \latin{et~al.}(2022)Lee, Hsieh, Zhang, and Shi]{Lee2022}
Lee,~C.-K.; Hsieh,~C.-Y.; Zhang,~S.; Shi,~L. Variational quantum simulation of
  chemical dynamics with quantum computers. \emph{Journal of Chemical Theory
  and Computation} \textbf{2022}, \emph{18}, 2105--2113\relax
\mciteBstWouldAddEndPuncttrue
\mciteSetBstMidEndSepPunct{\mcitedefaultmidpunct}
{\mcitedefaultendpunct}{\mcitedefaultseppunct}\relax
\EndOfBibitem
\bibitem[Miessen \latin{et~al.}(2021)Miessen, Ollitrault, and
  Tavernelli]{Miessen2021}
Miessen,~A.; Ollitrault,~P.~J.; Tavernelli,~I. Quantum algorithms for quantum
  dynamics: A performance study on the spin-boson model. \emph{Physical Review
  Research} \textbf{2021}, \emph{3}, 043212\relax
\mciteBstWouldAddEndPuncttrue
\mciteSetBstMidEndSepPunct{\mcitedefaultmidpunct}
{\mcitedefaultendpunct}{\mcitedefaultseppunct}\relax
\EndOfBibitem
\bibitem[Alghassi \latin{et~al.}(2022)Alghassi, Deshmukh, Ibrahim, Robles,
  Woerner, and Zoufal]{Alghassi2022}
Alghassi,~H.; Deshmukh,~A.; Ibrahim,~N.; Robles,~N.; Woerner,~S.; Zoufal,~C. A
  variational quantum algorithm for the Feynman-Kac formula. \emph{Quantum}
  \textbf{2022}, \emph{6}, 730\relax
\mciteBstWouldAddEndPuncttrue
\mciteSetBstMidEndSepPunct{\mcitedefaultmidpunct}
{\mcitedefaultendpunct}{\mcitedefaultseppunct}\relax
\EndOfBibitem
\bibitem[Nakaji \latin{et~al.}(2023)Nakaji, Endo, Matsuzaki, and
  Hakoshima]{Nakaji2023}
Nakaji,~K.; Endo,~S.; Matsuzaki,~Y.; Hakoshima,~H. Measurement optimization of
  variational quantum simulation by classical shadow and derandomization.
  \emph{Quantum} \textbf{2023}, \emph{7}, 995\relax
\mciteBstWouldAddEndPuncttrue
\mciteSetBstMidEndSepPunct{\mcitedefaultmidpunct}
{\mcitedefaultendpunct}{\mcitedefaultseppunct}\relax
\EndOfBibitem
\bibitem[Chen \latin{et~al.}(2024)Chen, Gomes, Niu, and de~Jong]{Chen2024}
Chen,~H.; Gomes,~N.; Niu,~S.; de~Jong,~W.~A. Adaptive variational simulation
  for open quantum system dynamics. \emph{Quantum} \textbf{2024},
  \emph{8}\relax
\mciteBstWouldAddEndPuncttrue
\mciteSetBstMidEndSepPunct{\mcitedefaultmidpunct}
{\mcitedefaultendpunct}{\mcitedefaultseppunct}\relax
\EndOfBibitem
\bibitem[Linteau \latin{et~al.}(2024)Linteau, Barison, Lindner, and
  Carleo]{Linteau2024}
Linteau,~D.; Barison,~S.; Lindner,~N.~H.; Carleo,~G. Adaptive projected
  variational quantum dynamics. \emph{Phys. Rev. Res.} \textbf{2024}, \emph{6},
  023130\relax
\mciteBstWouldAddEndPuncttrue
\mciteSetBstMidEndSepPunct{\mcitedefaultmidpunct}
{\mcitedefaultendpunct}{\mcitedefaultseppunct}\relax
\EndOfBibitem
\bibitem[Gulliksson \latin{et~al.}(2000)Gulliksson, Wedin, and
  Wei]{Gulliksson2000}
Gulliksson,~M.; Wedin,~P.-{\AA}.; Wei,~Y. Perturbation identities for
  regularized Tikhonov inverses and weighted pseudoinverses. \emph{BIT
  Numerical Mathematics} \textbf{2000}, \emph{40}, 513--523\relax
\mciteBstWouldAddEndPuncttrue
\mciteSetBstMidEndSepPunct{\mcitedefaultmidpunct}
{\mcitedefaultendpunct}{\mcitedefaultseppunct}\relax
\EndOfBibitem
\bibitem[Dogra \latin{et~al.}(2021)Dogra, Melnikov, and Paraoanu]{Dogra2021}
Dogra,~S.; Melnikov,~A.~A.; Paraoanu,~G.~S. Quantum simulation of parity–time
  symmetry breaking with a superconducting quantum processor.
  \emph{Communications Physics} \textbf{2021}, \emph{4}\relax
\mciteBstWouldAddEndPuncttrue
\mciteSetBstMidEndSepPunct{\mcitedefaultmidpunct}
{\mcitedefaultendpunct}{\mcitedefaultseppunct}\relax
\EndOfBibitem
\bibitem[Fenna and Matthews(1975)Fenna, and Matthews]{Fenna1975}
Fenna,~R.~E.; Matthews,~B.~W. Chlorophyll arrangement in a bacteriochlorophyll
  protein from Chlorohium limicola. \emph{Nature} \textbf{1975}, \emph{258},
  1476--4687\relax
\mciteBstWouldAddEndPuncttrue
\mciteSetBstMidEndSepPunct{\mcitedefaultmidpunct}
{\mcitedefaultendpunct}{\mcitedefaultseppunct}\relax
\EndOfBibitem
\bibitem[Ishizaki and Fleming(2009)Ishizaki, and Fleming]{Ishizaki2009}
Ishizaki,~A.; Fleming,~G. Unified treatment of quantum coherent and incoherent
  hopping dynamics in electronic energy transfer: Reduced hierarchy equation
  approach. \emph{The Journal of Chemical Physics} \textbf{2009}, \emph{130},
  234111\relax
\mciteBstWouldAddEndPuncttrue
\mciteSetBstMidEndSepPunct{\mcitedefaultmidpunct}
{\mcitedefaultendpunct}{\mcitedefaultseppunct}\relax
\EndOfBibitem
\bibitem[Zhu \latin{et~al.}(2012)Zhu, \latin{et~al.} others]{Zhu2012}
Zhu,~J., \latin{et~al.}  Multipartite quantum entanglement evolution in
  photosynthetic complexes. \emph{The Journal of Chemical Physics}
  \textbf{2012}, \emph{137}, 074112\relax
\mciteBstWouldAddEndPuncttrue
\mciteSetBstMidEndSepPunct{\mcitedefaultmidpunct}
{\mcitedefaultendpunct}{\mcitedefaultseppunct}\relax
\EndOfBibitem
\bibitem[Thyrhaug \latin{et~al.}(2018)Thyrhaug, \latin{et~al.}
  others]{Thyrhaug2018}
Thyrhaug,~E., \latin{et~al.}  Identification and characterization of diverse
  coherences in the Fenna–Matthews–Olson complex. \emph{Nature Chemistry}
  \textbf{2018}, \emph{10}, 780--786\relax
\mciteBstWouldAddEndPuncttrue
\mciteSetBstMidEndSepPunct{\mcitedefaultmidpunct}
{\mcitedefaultendpunct}{\mcitedefaultseppunct}\relax
\EndOfBibitem
\bibitem[Irgen-Gioro \latin{et~al.}(2019)Irgen-Gioro, \latin{et~al.}
  others]{Irgengioro2019}
Irgen-Gioro,~S., \latin{et~al.}  Electronic coherence lifetimes of the
  Fenna–Matthews–Olson complex and light harvesting complex II.
  \emph{Chemical Science} \textbf{2019}, \emph{10}, 10503--10509\relax
\mciteBstWouldAddEndPuncttrue
\mciteSetBstMidEndSepPunct{\mcitedefaultmidpunct}
{\mcitedefaultendpunct}{\mcitedefaultseppunct}\relax
\EndOfBibitem
\bibitem[Oh \latin{et~al.}(2019)Oh, Coker, and Hutchinson]{Oh2019}
Oh,~S.; Coker,~D.; Hutchinson,~D. Optimization of energy transport in the
  Fenna-Matthews-Olson complex via site-varying pigment-protein interactions.
  \emph{The Journal of Chemical Physics} \textbf{2019}, \emph{150},
  085102\relax
\mciteBstWouldAddEndPuncttrue
\mciteSetBstMidEndSepPunct{\mcitedefaultmidpunct}
{\mcitedefaultendpunct}{\mcitedefaultseppunct}\relax
\EndOfBibitem
\bibitem[Kim \latin{et~al.}(2020)Kim, \latin{et~al.} others]{Kim2020}
Kim,~Y., \latin{et~al.}  Predictive First-Principles Modeling of a
  Photosynthetic Antenna Protein: The Fenna–Matthews–Olson Complex.
  \emph{The Journal of Physical Chemistry Letters} \textbf{2020}, \emph{11},
  1636--1643\relax
\mciteBstWouldAddEndPuncttrue
\mciteSetBstMidEndSepPunct{\mcitedefaultmidpunct}
{\mcitedefaultendpunct}{\mcitedefaultseppunct}\relax
\EndOfBibitem
\bibitem[Suzuki \latin{et~al.}(2020)Suzuki, \latin{et~al.} others]{Suzuki2020}
Suzuki,~Y., \latin{et~al.}  Comparative study on model parameter evaluations
  for the energy transfer dynamics in Fenna–Matthews–Olson complex.
  \emph{Chemical Physics} \textbf{2020}, \emph{539}, 110903\relax
\mciteBstWouldAddEndPuncttrue
\mciteSetBstMidEndSepPunct{\mcitedefaultmidpunct}
{\mcitedefaultendpunct}{\mcitedefaultseppunct}\relax
\EndOfBibitem
\bibitem[Engel \latin{et~al.}(2007)Engel, Calhoun, Read, Ahn, Man{\v{c}}al,
  Cheng, Blankenship, and Fleming]{Engel2007}
Engel,~G.~S.; Calhoun,~T.~R.; Read,~E.~L.; Ahn,~T.-K.; Man{\v{c}}al,~T.;
  Cheng,~Y.-C.; Blankenship,~R.~E.; Fleming,~G.~R. {Evidence for wavelike
  energy transfer through quantum coherence in photosynthetic systems}.
  \emph{Nature} \textbf{2007}, \emph{446}\relax
\mciteBstWouldAddEndPuncttrue
\mciteSetBstMidEndSepPunct{\mcitedefaultmidpunct}
{\mcitedefaultendpunct}{\mcitedefaultseppunct}\relax
\EndOfBibitem
\bibitem[Lee \latin{et~al.}(2007)Lee, Cheng, and Fleming]{Lee2007}
Lee,~H.; Cheng,~Y.-C.; Fleming,~G.~R. {Coherence Dynamics in Photosynthesis:
  Protein Protection of Excitonic Coherence}. \emph{Science} \textbf{2007},
  \emph{316}, 1462--1465\relax
\mciteBstWouldAddEndPuncttrue
\mciteSetBstMidEndSepPunct{\mcitedefaultmidpunct}
{\mcitedefaultendpunct}{\mcitedefaultseppunct}\relax
\EndOfBibitem
\bibitem[Sension(2007)]{Sension2007}
Sension,~R. Quantum path to photosynthesis. \emph{Nature} \textbf{2007},
  \emph{446}, 740--741\relax
\mciteBstWouldAddEndPuncttrue
\mciteSetBstMidEndSepPunct{\mcitedefaultmidpunct}
{\mcitedefaultendpunct}{\mcitedefaultseppunct}\relax
\EndOfBibitem
\bibitem[Barroso-Flores(2017)]{Barroso-Flores2017}
Barroso-Flores,~J. Evolution of the Fenna–Matthews–Olson Complex and Its
  Quantum Coherence Features. Which Led the Way? \emph{ACS Central Science}
  \textbf{2017}, \emph{3}, 1061--1062\relax
\mciteBstWouldAddEndPuncttrue
\mciteSetBstMidEndSepPunct{\mcitedefaultmidpunct}
{\mcitedefaultendpunct}{\mcitedefaultseppunct}\relax
\EndOfBibitem
\bibitem[Hu \latin{et~al.}(2018)Hu, amd Fahhad H.~Alharbi, and Kais]{Hu2018a}
Hu,~Z.; amd Fahhad H.~Alharbi,~G. S.~E.; Kais,~S. Dark states and
  delocalization: Competing effects of quantum coherence on the efficiency of
  light harvesting systems. \emph{The Journal of Chemical Physics}
  \textbf{2018}, \emph{148}, 064304\relax
\mciteBstWouldAddEndPuncttrue
\mciteSetBstMidEndSepPunct{\mcitedefaultmidpunct}
{\mcitedefaultendpunct}{\mcitedefaultseppunct}\relax
\EndOfBibitem
\bibitem[Hu \latin{et~al.}(2018)Hu, Engel, and Kais]{Hu2018b}
Hu,~Z.; Engel,~G.; Kais,~S. Double-excitation manifold's effect on exciton
  transfer dynamics and the efficiency of coherent light harvesting.
  \emph{Physical Chemistry Chemical Physics} \textbf{2018}, \emph{20},
  30032--30040\relax
\mciteBstWouldAddEndPuncttrue
\mciteSetBstMidEndSepPunct{\mcitedefaultmidpunct}
{\mcitedefaultendpunct}{\mcitedefaultseppunct}\relax
\EndOfBibitem
\bibitem[Skochdopole and Mazziotti(2011)Skochdopole, and
  Mazziotti]{Skochdopole2011}
Skochdopole,~N.; Mazziotti,~D. Functional Subsystems and Quantum Redundancy in
  Photosynthetic Light Harvesting. \emph{The Journal of Physical Chemistry
  Letters} \textbf{2011}, \emph{2}, 2989--2993\relax
\mciteBstWouldAddEndPuncttrue
\mciteSetBstMidEndSepPunct{\mcitedefaultmidpunct}
{\mcitedefaultendpunct}{\mcitedefaultseppunct}\relax
\EndOfBibitem
\bibitem[Avdic \latin{et~al.}(2023)Avdic, Sager-Smith, Ghosh, Wedig, Higgins,
  Engel, and Mazziotti]{Avdic2023}
Avdic,~I.; Sager-Smith,~L.~M.; Ghosh,~I.; Wedig,~O.~C.; Higgins,~J.~S.;
  Engel,~G.~S.; Mazziotti,~D.~A. Quantum sensing using multiqubit quantum
  systems and the Pauli polytope. \emph{Phys. Rev. Res.} \textbf{2023},
  \emph{5}, 043097\relax
\mciteBstWouldAddEndPuncttrue
\mciteSetBstMidEndSepPunct{\mcitedefaultmidpunct}
{\mcitedefaultendpunct}{\mcitedefaultseppunct}\relax
\EndOfBibitem
\bibitem[Schouten \latin{et~al.}(2023)Schouten, Sager-Smith, and
  Mazziotti]{Schouten2023}
Schouten,~A.~O.; Sager-Smith,~L.~M.; Mazziotti,~D.~A. {Exciton-Condensate-Like
  Amplification of Energy Transport in Light Harvesting}. \emph{PRX Energy}
  \textbf{2023}, \emph{2}, 023002\relax
\mciteBstWouldAddEndPuncttrue
\mciteSetBstMidEndSepPunct{\mcitedefaultmidpunct}
{\mcitedefaultendpunct}{\mcitedefaultseppunct}\relax
\EndOfBibitem
\bibitem[Seneviratne \latin{et~al.}(2024)Seneviratne, Walters, and
  Wang]{Seneviratne2024}
Seneviratne,~A.; Walters,~P.~L.; Wang,~F. Exact Non-Markovian Quantum Dynamics
  on the NISQ Device Using Kraus Operators. \emph{ACS Omega} \textbf{2024},
  \emph{9}, 9666--9675\relax
\mciteBstWouldAddEndPuncttrue
\mciteSetBstMidEndSepPunct{\mcitedefaultmidpunct}
{\mcitedefaultendpunct}{\mcitedefaultseppunct}\relax
\EndOfBibitem
\bibitem[Welch \latin{et~al.}(2014)Welch, Greenbaum, Mostame, and
  Aspuru-Guzik]{Welch2014}
Welch,~J.; Greenbaum,~D.; Mostame,~S.; Aspuru-Guzik,~A. Efficient quantum
  circuits for diagonal unitaries without ancillas. \emph{New Journal of
  Physics} \textbf{2014}, \emph{16}, 033040\relax
\mciteBstWouldAddEndPuncttrue
\mciteSetBstMidEndSepPunct{\mcitedefaultmidpunct}
{\mcitedefaultendpunct}{\mcitedefaultseppunct}\relax
\EndOfBibitem
\bibitem[Ritz \latin{et~al.}(2004)Ritz, \latin{et~al.} others]{Ritz2004}
Ritz,~T., \latin{et~al.}  Resonance effects indicate a radical-pair mechanism
  for avian magnetic compass. \emph{Nature} \textbf{2004}, \emph{429},
  177--180\relax
\mciteBstWouldAddEndPuncttrue
\mciteSetBstMidEndSepPunct{\mcitedefaultmidpunct}
{\mcitedefaultendpunct}{\mcitedefaultseppunct}\relax
\EndOfBibitem
\bibitem[Rodgers and Hore(2009)Rodgers, and Hore]{Rodgers2009}
Rodgers,~C.; Hore,~P. Chemical magnetoreception in birds: The radical pair
  mechanism. \emph{Proceedings of the National Academy of Sciences}
  \textbf{2009}, \emph{106}, 353--360\relax
\mciteBstWouldAddEndPuncttrue
\mciteSetBstMidEndSepPunct{\mcitedefaultmidpunct}
{\mcitedefaultendpunct}{\mcitedefaultseppunct}\relax
\EndOfBibitem
\bibitem[Gauger \latin{et~al.}(2011)Gauger, \latin{et~al.} others]{Gauger2011}
Gauger,~E., \latin{et~al.}  Sustained Quantum Coherence and Entanglement in the
  Avian Compass. \emph{Physical Review Letters} \textbf{2011}, \emph{106},
  040503\relax
\mciteBstWouldAddEndPuncttrue
\mciteSetBstMidEndSepPunct{\mcitedefaultmidpunct}
{\mcitedefaultendpunct}{\mcitedefaultseppunct}\relax
\EndOfBibitem
\bibitem[Bender and Boettcher(1998)Bender, and Boettcher]{Bender1998}
Bender,~C.~M.; Boettcher,~S. Real Spectra in Non-Hermitian Hamiltonians Having
  PT Symmetry. \emph{Physical Review Letters} \textbf{1998}, \emph{80},
  5243–5246\relax
\mciteBstWouldAddEndPuncttrue
\mciteSetBstMidEndSepPunct{\mcitedefaultmidpunct}
{\mcitedefaultendpunct}{\mcitedefaultseppunct}\relax
\EndOfBibitem
\bibitem[El-Ganainy \latin{et~al.}(2018)El-Ganainy, Makris, Khajavikhan,
  Musslimani, Rotter, and Christodoulides]{ElGanainy2018}
El-Ganainy,~R.; Makris,~K.~G.; Khajavikhan,~M.; Musslimani,~Z.~H.; Rotter,~S.;
  Christodoulides,~D.~N. Non-Hermitian physics and PT symmetry. \emph{Nature
  Physics} \textbf{2018}, \emph{14}, 11–19\relax
\mciteBstWouldAddEndPuncttrue
\mciteSetBstMidEndSepPunct{\mcitedefaultmidpunct}
{\mcitedefaultendpunct}{\mcitedefaultseppunct}\relax
\EndOfBibitem
\bibitem[Wang \latin{et~al.}(2009)Wang, Chong, Joannopoulos, and
  Soljačić]{Wang2009}
Wang,~Z.; Chong,~Y.; Joannopoulos,~J.~D.; Soljačić,~M. Observation of
  unidirectional backscattering-immune topological electromagnetic states.
  \emph{Nature} \textbf{2009}, \emph{461}, 772–775\relax
\mciteBstWouldAddEndPuncttrue
\mciteSetBstMidEndSepPunct{\mcitedefaultmidpunct}
{\mcitedefaultendpunct}{\mcitedefaultseppunct}\relax
\EndOfBibitem
\bibitem[Patil \latin{et~al.}(2022)Patil, H\"{o}ller, Henry, Guria, Zhang,
  Jiang, Kralj, Read, and Harris]{Patil2022}
Patil,~Y. S.~S.; H\"{o}ller,~J.; Henry,~P.~A.; Guria,~C.; Zhang,~Y.; Jiang,~L.;
  Kralj,~N.; Read,~N.; Harris,~J. G.~E. Measuring the knot of non-Hermitian
  degeneracies and non-commuting braids. \emph{Nature} \textbf{2022},
  \emph{607}, 271–275\relax
\mciteBstWouldAddEndPuncttrue
\mciteSetBstMidEndSepPunct{\mcitedefaultmidpunct}
{\mcitedefaultendpunct}{\mcitedefaultseppunct}\relax
\EndOfBibitem
\bibitem[Wu \latin{et~al.}(2019)Wu, Liu, Geng, Song, Ye, Duan, Rong, and
  Du]{Wu2019}
Wu,~Y.; Liu,~W.; Geng,~J.; Song,~X.; Ye,~X.; Duan,~C.-K.; Rong,~X.; Du,~J.
  Observation of parity-time symmetry breaking in a single-spin system.
  \emph{Science} \textbf{2019}, \emph{364}, 878–880\relax
\mciteBstWouldAddEndPuncttrue
\mciteSetBstMidEndSepPunct{\mcitedefaultmidpunct}
{\mcitedefaultendpunct}{\mcitedefaultseppunct}\relax
\EndOfBibitem
\bibitem[Naghiloo \latin{et~al.}(2019)Naghiloo, Abbasi, Joglekar, and
  Murch]{Naghiloo2019}
Naghiloo,~M.; Abbasi,~M.; Joglekar,~Y.~N.; Murch,~K.~W. Quantum state
  tomography across the exceptional point in a single dissipative qubit.
  \emph{Nature Physics} \textbf{2019}, \emph{15}, 1232–1236\relax
\mciteBstWouldAddEndPuncttrue
\mciteSetBstMidEndSepPunct{\mcitedefaultmidpunct}
{\mcitedefaultendpunct}{\mcitedefaultseppunct}\relax
\EndOfBibitem
\bibitem[Kazmina \latin{et~al.}(2024)Kazmina, Zalivako, Borisenko, Nemkov,
  Nikolaeva, Simakov, Kuznetsova, Egorova, Galstyan, Semenin, Korolkov,
  Moskalenko, Abramov, Besedin, Kalacheva, Lubsanov, Bolgar, Kiktenko,
  Khabarova, Galda, Semerikov, Kolachevsky, Maleeva, and Fedorov]{Kazmina2024}
Kazmina,~A.~S. \latin{et~al.}  Demonstration of a parity-time-symmetry-breaking
  phase transition using superconducting and trapped-ion qutrits.
  \emph{Physical Review A} \textbf{2024}, \emph{109}\relax
\mciteBstWouldAddEndPuncttrue
\mciteSetBstMidEndSepPunct{\mcitedefaultmidpunct}
{\mcitedefaultendpunct}{\mcitedefaultseppunct}\relax
\EndOfBibitem
\end{mcitethebibliography}

\clearpage

\begin{figure}[h!]
    \centering
    \includegraphics[width = 0.85\textwidth]{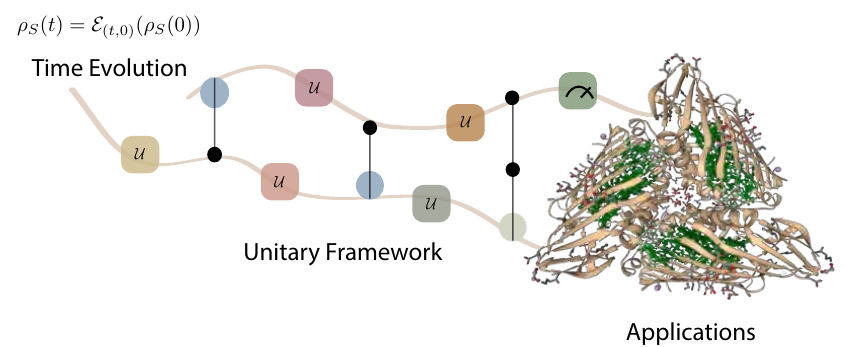}
    \caption{TOC Graphics}
\end{figure}

\end{document}